\documentclass[prd,amsmath,amssymb,showpacs]{revtex4}

\usepackage{mathrsfs,latexsym,bm}
\usepackage[dvips]{graphics}
\usepackage{pst-all}

\renewcommand{\square}{\vrule height 1.5ex width 1.2ex depth -.1ex }

\newcommand{\II}{\leavevmode\hbox{\rm{\small1\kern-3.8pt\normalsize1}}}

\newcommand{\RR}{{\mathbb R}}

\newcommand{\NN}{{\mathbb N}}
\newcommand{\TT}{{\mathbb T}}
\newcommand{\ZZ}{{\mathbb Z}}

\newcommand{\CoinX}[1]{C_0^\infty({#1})}
\newcommand{\tCoinX}[1]{\widetilde{C}_0^\infty({#1})}

\newtheorem{Thm}{Theorem}[section]
\newtheorem{Def}[Thm]{Definition}
\newtheorem{Lem}[Thm]{Lemma}
\newtheorem{Prop}[Thm]{Proposition}
\newtheorem{Cor}[Thm]{Corollary}

\newcommand{\DD}{{\mathscr D}}
\newcommand{\EE}{{\mathscr E}}

\newcommand{\FF}{{\mathscr F}}

\newcommand{\OO}{{\cal O}}

\newcommand{\QQ}{{\mathcal Q}}

\newcommand{\Ss}{{\mathcal S}}

\newcommand{\gb}{{\boldsymbol{g}}}

\newcommand{\xb}{{\boldsymbol{x}}}

\newcommand{\Af}{{\mathfrak A}}

\newcommand{\Dal}{\Box}

\newcommand{\supp}{{\rm supp}\,}

\newcommand{\ip}[2]{{\langle #1\mid #2\rangle}}

\newcommand{\stack}[2]{\substack{#1 \\ #2}}
\newcommand{\expt}[2]{{\langle #1 \rangle}_{#2}}

\newcommand{\eb}{{\boldsymbol{e}}}
\newcommand{\Cb}{{\boldsymbol{C}}}
\newcommand{\Mb}{{\boldsymbol{M}}}
\newcommand{\Nb}{{\boldsymbol{N}}}
\newcommand{\Rb}{{\boldsymbol{R}}}
\newcommand{\Ub}{{\boldsymbol{U}}}
\newcommand{\Ts}{{\sf T}}

\newcommand{\es}{{\sf e}}
\newcommand{\fs}{{\sf f}}
\newcommand{\ks}{{\sf k}}
\newcommand{\ts}{{\sf t}}
\newcommand{\us}{{\sf u}}
\newcommand{\vs}{{\sf v}}
\newcommand{\xis}{{\sf \xi}}

\DeclareMathOperator{\csch}{cosech}

\newcommand{\dvol}{d{\rm vol}}

\begin{document}


\title{Quantum Energy Inequalities and local covariance I: Globally
hyperbolic spacetimes}
\author{Christopher J. Fewster}
\email{cjf3@york.ac.uk}
\affiliation{Department of Mathematics, University of York, %
Heslington, York, YO10 5DD, UK}
\author{Michael J. Pfenning}
\email{Michael.Pfenning@usma.edu}
\affiliation{Department of Physics, United States Military Academy,%
 West Point, New York, 10996-1790, USA}

\date{\today}

\begin{abstract}
We begin a systematic study of Quantum Energy
Inequalities (QEIs) in relation to local covariance. We define notions of locally
covariant QEIs of both `absolute' and `difference' types and show that
existing QEIs satisfy these conditions. Local covariance permits us to
place constraints on the renormalised stress-energy tensor in
one spacetime using QEIs derived in another, in subregions where the two
spacetimes are isometric. This is of particular utility where one of the two
spacetimes exhibits a high degree of symmetry and the QEIs are available
in simple closed form. Various general applications are presented,
including {\em a priori} constraints (depending only on geometric
quantities) on the ground-state energy density
in a static spacetime containing locally Minkowskian regions. 
In addition, we present a number of
concrete calculations in both two and four dimensions which demonstrate
the consistency of our bounds with various known ground- and thermal state
energy densities. Examples considered include the Rindler and Misner
spacetimes, and spacetimes with toroidal spatial sections.
In this paper we confine the discussion to globally hyperbolic
spacetimes; subsequent papers will also discuss spacetimes with
boundary and other related issues.
\end{abstract}

\maketitle


\section{Introduction}

\label{sec:int_summary}

Over the past 30 years, much effort has been devoted to calculations of the 
renormalised
stress-energy tensor in ground states of quantum fields on stationary
background spacetimes. Many analogous calculations have been made in
flat spacetime equipped with reflecting boundaries, in connection with
the Casimir effect. However, it would be fair to say that only limited
qualititative insight has been gained. For example, the energy density
is sometimes positive, and sometimes negative and there is no known way of
predicting the sign in any general situations without performing the
full calculations~\footnote{This point has often been emphasised by
L.H.\ Ford.} (see, however,~\cite{Ken&Kli06} for a situation where the
sign can be predicted).
At least analytically, these calculations are
restricted to cases exhibiting a high degree of symmetry. The aim of
this paper, and a companion paper~\cite{cas_b}, is to point out
that there are situations in which one may gain some qualitative insight
into the possible magnitude of the stress-energy tensor
based on simple geometric considerations. 

The situation we study in this paper arises when a spacetime contains a 
subspacetime
which is isometric to (a subspacetime of) another spacetime, which will
usually have nontrivial symmetries. By using quantum energy
inequalities (QEIs) together with the locality properties of quantum field
theory, we are then able to use information about the second (symmetric)
spacetime to yield information about the stress-energy tensor of states 
on the first spacetime
(which need have no global symmetries) in the region where the
isometry holds. We will work on globally hyperbolic spacetimes in
this paper, deferring the issue of spacetimes with boundary to a
companion paper~\cite{cas_b}. As well as setting out the theory
behind the method, we will demonstrate it in several locally Minkowskian
spacetimes. Marecki~\cite{Marecki05} has also illustrated our approach, by
considering the case of spacetimes locally isometric to portions of exterior
Schwarzschild.  Also begun here for the free massless scalar field
is a similar discussion for conformally related 
regions of two-dimensional spacetimes. In a separate paper we will
extend this to the generalised Maxwell field in higher dimensional 
manifolds related by conformal diffeomorphisms.

To be more specific, consider a globally hyperbolic spacetime $\Nb$,
consisting of a manifold of dimension $d\ge 2$, a Lorentzian metric
with signature $+-\cdots-$, and choices of
orientation and time-orientation (which, together, are required to
fulfill the demands of global hyperbolicity) \footnote{To be more
precise, the spacetime manifold is required to be connected, smooth, 
Hausdorff, and
paracompact. The spacetime $\Nb$ is globally hyperbolic if it contains a
Cauchy surface, i.e., a subset intersected exactly once by
every inextendible timelike curve~\cite{ONeill83}. The globally hyperbolic
spacetimes are the most general class of spacetimes on which quantum
fields are typically formulated, but one should be aware that manifolds
with boundary are not included.}. Suppose an open subset of $\Nb$, 
when equipped with the metric and (time-)orientation inherited from
$\Nb$, is a globally hyperbolic spacetime $\Nb'$ in its own right. If, moreover,
any causal curve in $\Nb$ whose endpoints lie in $\Nb'$ is contained 
completely in $\Nb'$, then we will call $\Nb'$ a {\em causally embedded
globally hyperbolic subspacetime} (c.e.g.h.s.) of $\Nb$. Our main
interest will be in the situation where a c.e.g.h.s.\ $\Nb'$ of $\Nb$ is
isometric to a c.e.g.h.s.\ $\Mb'$ of a second globally hyperbolic 
spacetime $\Mb$, with the isometry also respecting the
(time-)orientation. (We speak of a {\em causal isometry} 
in this case.) By the principle of locality, we expect that
any experiment conducted within $\Nb'$ should have the
same results as the same experiment [i.e., its isometric image] conducted in
$\Mb'$. No observer in $\Nb'$ should be able to discern, by such local
experiments that she does not, in fact, inhabit $\Mb'$; in particular, 
energy densities in $\Nb'$ should be
subject to the same QEIs as those in $\Mb'$. We will demonstrate
explicitly that these expectations are met by the QEIs we employ. 

\begin{figure}
\begin{center}
\psset{xunit=3mm,yunit=3mm,runit=3mm}
\begin{pspicture}(-15,1)(15,15)

\newgray{palegray}{.9}

\pscustom[linewidth=1.5pt]{
   \psline{-}(-14,3)(-14,14)
   \psline{-}(-14,14)(-3,14)
   \psline{-}(-3,14)(-3,3)
   \psline{-}(-3,3)(-14,3)
   \closepath
}
\rput*(-8.5,2){$\bm M$}

\pscustom[linewidth=1.5pt,fillstyle=solid,fillcolor=palegray]{
   \psline{-}(-4.5,8.5)(-8.5,12.5)(-12.5,8.5)(-8.5,4.5)(-4.5,8.5)
   \closepath
}
\rput(-8.5,8.5){${\bm M}'$}

\pscustom[linewidth=1.5pt]{
   \psbezier{-}(3,3)(1,5)(4,10)(2,14)
   \psbezier{-}(2,14)(4,12.5)(12,16)(14,14)
   \psbezier{-}(14,14)(13,13)(15,6)(13,3)
   \psbezier{-}(13,3)(10,5)(4,2.5)(3,3)
   \closepath
}
\rput*(8,2){$\bm N$}

\pscustom[linewidth=1.5pt,fillstyle=solid,fillcolor=palegray]{
   \psbezier{-}(8,4)(7,6)(4.5,7)(4,8.5)
   \psbezier{-}(4,8.5)(5,9)(7,13)(8,13)
   \psbezier{-}(8,13)(10,12)(11,9)(12,9)
   \psbezier{-}(12,9)(11,7)(9,5)(8,4)
   \closepath
}
\rput(10.5,8.5){${\bm N}'$}

\psbezier[linewidth=2.5pt]{->}(8,5)(7,8)(9,10)(8,12.5)
\rput(7,8.75){$\gamma$}

\psbezier[linewidth=4pt]{->}(-5,9)(-1,10)(0,10)(4.6,9)
\rput*(-0.2,11){$\psi$}

\end{pspicture} 
\end{center}
\caption{Illustration for Example 1: The curve $\gamma$ in $\Nb$ is enclosed in
a causally embedded globally hyperbolic subspacetime ${\bm N}'$ which is causally isometric to a
causally embedded globally hyperbolic subspacetime $\Mb'$
of four-dimensional Minkowski space $\Mb$ under $\psi:\Mb'\to\Nb'$.}
\label{fig:Mink_embed}
\end{figure}
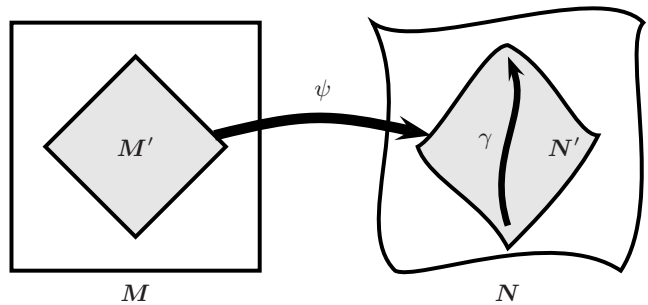

Among our results are the following, which we state for the case of a
Klein--Gordon field of mass $m\ge 0$ in four dimensions: 

{\noindent\em Example 1:} Suppose a timelike {\em geodesic} segment 
$\gamma$ of proper duration $\tau_0$ in a
globally hyperbolic spacetime $\Nb$ can be enclosed in a c.e.g.h.s.\
$\Nb'$ which is causally isometric to a c.e.g.h.s.\ of four-dimensional Minkowski
space as shown in Fig.~\ref{fig:Mink_embed}. Then any state $\omega$ of the 
Klein--Gordon field (of mass $m\ge 0$) on $\Nb$ obeys
\begin{equation}
\sup_\gamma \langle T_{ab}u^a u^b\rangle_\omega \ge -\frac{C_4}{\tau_0^4}
\end{equation}
where the constant $C_4= 3.169858\ldots$ (if $m>0$, one may obtain even more rapid
decay). 

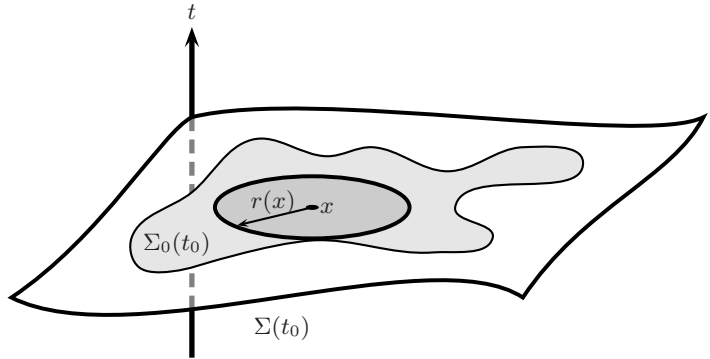
\begin{figure}
\begin{center}
\psset{xunit=4mm,yunit=4mm,runit=4mm}
\begin{pspicture}(-7,0)(20,11)

\newgray{palegray}{.9}
\newgray{pmgray}{.8}

\psline[linewidth=2pt]{-}(0,0)(0,1.6)
\psline[linewidth=2pt,linestyle=dashed,linecolor=gray]{-}(0,1.6)(0,8)
\psline[linewidth=2pt]{->}(0,8)(0,11)
\rput(0,11.5){$t$}


\pscustom[linewidth=1.5pt]{
   \psbezier{-}(0,8)(4,9)(15,7)(17,8)
   \psbezier{-}(17,8)(16,6)(13,5)(11,2)
   \psbezier{-}(11,2)(8,4)(-2,0)(-6,2)
   \psbezier{-}(-6,2)(-3,4)(-1,7.5)(0,8)
   \closepath
}

\pscurve[showpoints=false,fillstyle=solid,fillcolor=palegray]{-}(0,3)(4,3.9)(8,3.5)(10,4)(8.75,5)(13,6.5)(11,7)%
(8.5,6.2)(6,7)(4.5,6.7)(2,7.25)(0,5.5)(-2,4)(-1.75,3)(0,3)

\psellipse[linewidth=1.5pt,fillstyle=solid,fillcolor=pmgray](4,5)(3.3,1.1)
\psellipse[linewidth=0pt,fillstyle=solid,fillcolor=black](4,5)(.2,.066)
\psline{->}(4,5)(1.5,4.4)
\rput(4.5,5){$x$}
\rput(2.7,5.2){$r(x)$}

\rput(3,1){$\Sigma(t_0)$}
\rput(-.5,3.8){$\Sigma_0(t_0)$}

\end{pspicture} 
\end{center}
\caption{Diagram showing open ball about the point $x$ in Example 2.}
\label{fig:example2}
\end{figure}

{\noindent\em Example 2:} Suppose a globally hyperbolic spacetime $\Nb$ 
is stationary with respect to a timelike
Killing field $t^a$ and admits the smooth foliation into constant time surfaces
$\Nb\cong\RR\times\Sigma$. Suppose the metric takes the Minkowski form (w.r.t.
some coordinates) on $\RR\times\Sigma_0$ for some subset $\Sigma_0$ of
$\Sigma$ with nonempty interior. (We may suppose that $\Sigma_0$ has been
taken to be maximal.) For any $x$ in the interior of $\Sigma_0$, 
let $r(x)$ be the radius of the largest Euclidean $3$-ball
which can be isometrically embedded in $\Sigma_0$, centred on $x$, as in Fig.~\ref{fig:example2}.
Then any stationary Hadamard state~\footnote{By a stationary state, we
mean one whose $n$-point functions are invariant under translations
along the Killing flow:
$w_n(\psi_t(x_1),\ldots,\psi_t(x_n))=w_n(x_1,\ldots,x_n)$, where
$\psi_t$ is the group of isometries associated with the Killing field.} $\omega_{\Nb}$ 
on $\Nb$ obeys the bound
\begin{equation}
\langle T_{ab}n^a n^b\rangle_{\omega_\Nb}(t,x)\ge -\frac{C_4}{(2r(x))^4}
\end{equation}
for any $x\in\Sigma_0$, where $n^a$ is the unit vector along $t^a$.

{\noindent\em Example 3:} Suppose $\gamma:\RR\to \Nb$ is a
uniformly accelerated trajectory (parametrised by proper time)
with proper acceleration $\alpha$,
and suppose $\gamma$ can be enclosed within a c.e.g.h.s. $\Nb'$ of $\Nb$
which is causally isometric to a c.e.g.h.s. of four-dimensional
Minkowski space. Then, for any Hadamard state $\omega$ on $\Nb$, and any
smooth compactly supported real-valued $g$, with $\int_{-\infty}^\infty g(\tau)^2\,d\tau=1$,
\begin{equation}
\liminf_{\tau_0\to\infty} \frac{1}{\tau_0}\int_\gamma \langle T_{ab}u^a u^b\rangle_\omega
g(\tau/\tau_0)^2\,d\tau
\ge -\frac{11\alpha^4}{480\pi^2}\,.
\end{equation}
Note the remarkable fact that the right-hand
side is precisely the expected energy density in the Rindler vacuum
state along the trajectory with constant proper accleration $\alpha$. In
particular, if the energy density in some state $\omega$ is constant
along $\gamma$, it must exceed or equal that of the Rindler vacuum. We emphasize
that our derivation does not involve the Rindler vacuum, but only the 
Minkowski vacuum state two-point function and the QEIs.

Variants of these results hold in other dimensions, and also for
other linear field equations such as the Maxwell and Proca fields
(which we will treat elsewhere).

To prepare for our main discussion, it will be useful to
make a few general remarks about quantum energy inequalities (QEIs),
also often called simply quantum inequalities (QIs). 
QEIs have been quite intensively developed over the past decade,
following Ford's much earlier insight~\cite{Ford78} that quantum
field theory might act to limit the magnitude and duration of negative
energy densities and/or fluxes, thereby preventing macroscopic
violations of the second law of thermodynamics (see~\cite{Fe&V03} for
rigorous links between QEIs and thermodynamical stability). Detailed reviews of
QEIs may be found in~\cite{Pfen98b,Fews05,Roma04}.

QEIs take various forms, but we will distinguish two basic types:
absolute QEIs and difference QEIs. An absolute QEI bound consists of a
set $\FF$ of {\em sampling tensors}, i.e., second rank contravariant tensor fields against which
the renormalised stress-energy tensor will be averaged, a class $\Ss$
of states of the theory (which may be chosen to have nice properties)
and a map $\QQ:\FF\to\RR$ such that
\begin{equation}
\int \langle T_{ab}\rangle_\omega f^{ab}\,\dvol \ge -\QQ(\fs)>-\infty
\end{equation}
for all states $\omega\in\Ss$~\footnote{It
would also be natural to demand that $\FF$ be convex (i.e., if $\fs_1$
and $\fs_2$ are in $\FF$ then so is $\lambda \fs_1+(1-\lambda)\fs_2$ for
all $\lambda\in[0,1]$, and for $\QQ$ to obey $\QQ(\lambda
\fs_1+(1-\lambda)\fs_2)\le \lambda \QQ(\fs_1)+(1-\lambda)\QQ(\fs_2)$,
but we shall not make these requirements.}. Here $T_{ab}$ is the renormalised
stress-energy tensor defined in a manner compatible with Wald's axioms
\cite{Wald_QFT}, and we have adopted the convention that the same tensor may be
written $\fs$ (without indices) or $f^{ab}$ (with). We will
permit $\FF$ to include tensors singularly supported on timelike curves
or other submanifolds of spacetime, so, for example, we can treat
worldline averages such as
\begin{equation}
\int_\gamma \langle T_{ab}\rangle_\omega f^{ab}\,d\tau = \int_I \langle
T_{ab}(\gamma(\tau))\rangle_\omega u^a u^b g(\tau)^2\,d\tau
\label{eq:worldline_average}
\end{equation}
where $\gamma$ is a smooth timelike curve parametrised by an open interval
$I$ of proper time $\tau$, with velocity $u^a$, and for $g\in\CoinX{I}$.
[To be precise, $\FF$ is required to be a set of
compactly supported distributions on smooth rank two covariant test
tensor fields.]

Absolute QEIs are known [with explicit formulae for $\QQ$, and specific
$\FF$ and $\Ss$] for (a) the scalar field
of mass $m\ge 0$ in $d$-dimensional Minkowski
space~\cite{F&Ro95,F&Ro97,Fe&E98} (see also~\cite{Ford91} for $d\ge 2$),
(b) the massless scalar and Fermi fields in arbitrary two-dimensional globally hyperbolic
spacetimes~\cite{Flan97,Voll00,Flan02,Fews04}, 
(c) general (interacting) conformal field theories in two-dimensional
Minkowski space~\cite{Fe&Ho05}, (d) a variety of higher spin linear
fields in two- and four-dimensional Minkowski
space~\cite{F&Ro97,Pfen02,Pfen03,Fe&Mi03,Daws06,Yu&Wu04,HuLiZh06}. For the most
part only worldline bounds involving averages of the form
Eq.~\eqref{eq:worldline_average} have been studied; it has been found
that replacing $g$ by a scaled version $\tau_0^{-1/2}g(\tau/\tau_0)$ has the effect of sending
the QEI bound to zero as $\tau_0^{-d}$ (or faster, for massive fields) as $\tau_0\to
\infty$, where $d$ is the spacetime dimension.

Difference QEI bounds also involve the specification of $\FF$ and $\Ss$
as before, but now the bound sought takes the form
\begin{equation}
\int \left[\langle T_{ab}\rangle_\omega-\langle T_{ab}\rangle_{\omega_0}\right] f^{ab}
\,\dvol \ge -\QQ(\fs,\omega_0)>-\infty\,,
\label{eq:QIdiff}
\end{equation}
where $\omega_0$ is called the reference state. If the theory were represented in a Fock space
built on $\omega_0$ (when this is possible) the left-hand side would be an
average of the normal ordered stress-energy tensor. However it is not always
necessary to assume that $\omega$ and $\omega_0$ are represented in this way. Difference QEIs have
proved to be the easiest to establish in curved spacetimes, or where
boundaries are present. First developed in the case of (ultra)static
spacetimes with the (ultra)static ground state chosen as the reference
state $\omega_0$~\cite{Pfen97a,Pfen98a,Fe&T99,Pfen02}, they are now known for
scalar, spin-$1/2$, and
spin-$1$ fields in arbitrary globally hyperbolic
spacetimes~\cite{Fews00,Fe&V02,Pfen03}. In these
general results, $\Ss$ is the class of Hadamard states and the bounds
are sufficiently general that $\omega_0$ may be any element of $\Ss$,
so $\QQ$ becomes a function $\QQ:\FF\times\Ss\to \RR$. The general
results do not make use of a Hilbert space representation.

Clearly, difference and absolute QEIs are quite closely related. In
particular, Wald's fourth axiom requires $\langle
T_{ab}\rangle_{\omega_0}$ to vanish identically if $\omega_0$ is the
Minkowski vacuum, so difference QEIs become absolute in this case. [The
extension of this observation to locally Minkowskian spaces is a key idea
in this paper.] More generally, we may convert a difference QEI to an absolute
QEI by moving all the terms in $\omega_0$
onto the right-hand side. In cases where the
renormalised stress-energy tensor is known explicitly for the reference state, this is
perfectly satisfactory. However, there are two (related) drawbacks:
(i) there is no canonical choice of reference state $\omega_0$ in a
general spacetime (which might have no timelike Killing fields, for
example); (ii) one does not normally have available a closed form
expression for $\langle T_{ab}\rangle_{\omega_0}$ for {\em any} state on
a general spacetime, so the QEI bound becomes somewhat inexplicit. 
This weakens the power of QEIs to constrain exotic spacetime configurations such as
macroscopic traversable wormholes or `warp drive'. (On sufficiently small scales, one expects
that the absolute QEI bounds should strongly resemble those of Minkowski
space---as first argued in \cite{F&Ro96}, and proven in various
situations in \cite{Pfen98a} and~\cite{Fews04}---however 
one still needs to know the magnitude of $\langle
T_{ab}\rangle_{\omega_0}$ to know on what scales this approximation holds.)

The present paper and its companion represent first steps towards
absolute QEIs in more general spacetimes, starting with spacetimes
containing regions isometric to others where reference states are known. 
Work is under way on generally applicable absolute QEIs and will be
reported elsewhere; however we
expect the results and methods presented here to be of continuing interest, as they
reduce to very simple geometrical conditions. 

The paper is structured as follows. In Sec.~\ref{sec:QEIs_and_lc} we
give a brief introduction to some of the relevant notions of locally
covariant quantum field theory before defining locally covariant QEIs
and developing their simple properties in Sec.~\ref{sec:QEI_loc_cov}.
The following two subsections show how existing QEIs in the literature
may be expressed in the locally covariant framework, and address some
technical points along the way. In Sec.~\ref{sec:general_apps} we show
how local covariance permits {\em a priori} bounds to be placed on
energy densities in spacetimes with Minkowskian subspacetimes using geometric
data. The main technique here, in
addition to local covariance, is the conversion of QEIs to eigenvalue
problems, first introduced in~\cite{Fe&Te00}. These are applied
 in Sec.~\ref{sec:specific_apps} to
specific spacetime models where the energy densities of ground- and
thermal states are known, permitting comparison with our {\em a priori}
bounds. In some cases these bounds are saturated by the exact values.
After a summary, the appendices collect various results needed in the
main text.

\section{Quantum energy inequalities and local covariance}\label{sec:QEIs_and_lc}

\subsection{Geometrical preliminaries}

Suppose two globally hyperbolic spacetimes of the same dimension, $\Mb_1$ and
$\Mb_2$, are given (we denote the corresponding manifolds and metrics by
$M_i$, $\gb_i$ for $i=1,2$). An {\em isometric embedding} of $\Mb_1$ in $\Mb_2$ 
is a smooth map $\psi:M_1\to M_2$ which is a diffeomorphism of
$M_1$ onto its range $\psi(M_1)$ in $M_2$ and so that the pull-back
$\psi^*\gb_2$ is everywhere equal to $\gb_1$ on $M_1$. In 
local coordinates, 
\begin{equation}
g_{1\,ab}(x)=\frac{\partial y^{a'}}{\partial {x}^{a}}
\frac{\partial y^{b'}}{\partial {x}^{b}}g_{2\,a'b'}(y)
\end{equation}
should hold for all $x\in M_1$, where $y=\psi(x)$. We {\em do} require
that all of $M_1$ is mapped into $M_2$, but we do {\em not} require that
the image of $M_1$ under $\psi$ consists of the whole of $M_2$. 
There are two
possible choices of orientation and time orientation on $\psi(M_1)$:
that induced by $\psi$ from the (time-)orientation of $\Mb_1$, and that
inherited from $\Mb_2$. If these coincide and we have the further
property that every causal curve in $\Mb_2$ with endpoints in $\psi(M_1)$ lies entirely in
$\psi(M_1)$, then we say that $\psi$ is a {\em causal isometric embedding}.
An important class of examples arises where $\Mb_1$ is a causally
embedded globally hyperbolic subspacetime (c.e.g.h.s.) of $\Mb_2$ as
defined in Sec.~\ref{sec:int_summary}, in which case
$\psi$ is simply the identity map. It is also worth
mentioning an example of a non-causal embedding,
namely, the `helical strip' described by
Kay~\cite{Kay92}. In this example a long thin diamond region of two dimensional
Minkowski space is isometrically embedded in a `timelike cylinder' which
is the quotient of Minkowski space by a spacelike translation. The
wrapping is arranged so that points which are spacelike separated in the
original diamond are timelike separated in the geometry of the timelike
cylinder. The definition of a causal embedding is designed precisely to
ensure that the induced and inherited causal structures cannot
differ in this way.

\subsection{Local covariance}\label{sect:localcovariance}

The relevance of local covariance to quantum field theory on manifolds has long been
understood~\cite{Kay79,Dimo80} but has recently been put in a new setting
by Brunetti, Fredenhagen and Verch~\cite{BrFrVe03} (see
also~\cite{Verch01} and~\cite{BrPo&Ru05}) and related work of
Hollands and Wald (see, e.g.,~\cite{Ho&Wa01,Ho&Wa02}). This provides a very elegant and general
framework for local covariance in the language of category theory.
However, we will only need a few of the main ideas of this analysis
and will not describe the whole structure.

In this section, we will restrict ourselves to the Klein--Gordon field of mass $m\ge
0$, although similar comments can be made for the Dirac, Maxwell, and
Proca fields. There is a well-defined quantisation of the theory on any globally
hyperbolic spacetime $\Mb$, in terms of an algebra of observables $\Af_\Mb$ and a space
of Hadamard states $\Ss_\Mb$ which determine expectation values for
observables in $\Af_\Mb$. For the purposes of this section, it
suffices to know that $\Af_\Mb$ is generated by smeared field objects
$\Phi_\Mb(f)$ labelled by smooth, compactly supported test functions
$f\in\CoinX{\Mb}$, subject to relations expressing the field equation and
commutation relations, and the hermiticity of the field. (The structure
is given in detail in Appendix~\ref{appdx:Locally_Covar_QFT}.) The Hadamard 
states of the theory are those states on $\Af_\Mb$ whose two-point functions 
have singularities of the Hadamard
form, which at leading order are just those of the Minkowski vacuum
two-point function. More precisely~\cite{Ka&Wa}, on any causal normal neighbourhood
$\OO$ in $\Mb$ there is a sequence of bidistributions
$H_n$ so that (for any $n$) the two-point function of any Hadamard state
differs from $H_n$ on $\OO$ by a state-dependent function of class $C^n$. It is
of key importance that $H_n(x,x')$ is fixed entirely by the local metric and causal
structure, through the Hadamard recursion relations. 
Given a Hadamard state $\omega$, we may construct the expected renormalised
stress-energy tensor $\langle T_{ab}\rangle_\omega$ by the
point-splitting technique (see, e.g.,~\cite{Wald_QFT}): first
subtract $H_n$ from the two-point
function (for $n\ge 2$), then
apply appropriate derivatives before taking the points together again.
Next, one subtracts a term of the form $Qg_{ab}$, where $Q$ is locally
determined (and state-independent), 
in order to ensure that the resulting tensor is conserved and vanishes
in the Minkowski vacuum state. The tensor defined in this way obeys
Wald's axioms mentioned above; however, these axioms would also be
satisfied if one were to add a conserved local curvature term. Such
terms are sometimes described as undetermined or arbitrary; we take
the view, however, that they are part of the specification of the
theory, just as the mass and conformal coupling are, even though they
do not appear explicitly in the Lagrangian (a similar attitude is
expressed in~\cite{Fl&W96}). For simplicity, and because our main
applications will concern locally Minkowskian spacetimes, we will
assume that these terms are absent -- that is, we restrict to those
scalar particle species for which this is the case. 

The above structure is locally covariant in the following sense. 
Suppose a globally hyperbolic spacetime $\Mb$ is embedded in a
globally hyperbolic spacetimes $\Nb$ by a causal isometry $\psi$, and
let $\psi_*$ denote the push-forward map on test functions. That is, 
$\psi_*:\CoinX{\Mb}\to\CoinX{\Nb}$ is defined by 
\begin{equation}
(\psi_*f)(y)=\left\{\begin{array}{cl} f(\psi^{-1}(y)) & \hbox{if
$y=\psi(x)$ for some $x\in \Mb$}\\ 
0 & \hbox{otherwise.}
\end{array}\right.
\end{equation}
Then there is a natural mapping of the field on $\Mb$ to the field on
$\Nb$ given by $\Phi_\Mb(f)\mapsto \Phi_\Nb(\psi_*f)$; we also write
this as $\psi_*(\Phi_\Mb(f))=\Phi_\Nb(\psi_*f)$. Moreover, $\psi_*$ can
be extended to any element of $\Af_\Mb$, respecting the algebraic
relations and mapping the identity in $\Af_\Mb$ to the identity in
$\Af_\Nb$; technically, it is a unit-preserving injective $*$-homomorphism of
$\Af_\Mb$ into $\Af_\Nb$. 

On account of the correspondence $\Phi_\Mb(f)\mapsto \Phi_\Nb(\psi_*f)$,
we say that the field is covariant [the transformation goes `in the same
direction' as $\psi$; see the remarks below on the
underlying category theory at the end of
Appendix~\ref{appdx:Locally_Covar_QFT}]. By contrast, the state
spaces transform in a contravariant way [in the `opposite direction' to $\psi$]:
for any state $\omega$ on
$\Af_\Nb$ there is a pulled-back state, which we denote $\psi^*\omega$,
on $\Af_\Mb$, so that the expectation values of $A\in\Af_\Mb$ and
$\psi_* A\in\Af_\Nb$ are related by
\begin{equation}
\langle A\rangle_{\psi^*\omega} = \langle \psi_*A \rangle_\omega\,.
\label{eq:exp_under_pullback}
\end{equation}
The use of pull-back notation may be justified by the observation that
Eq.~\eqref{eq:exp_under_pullback} entails that the $n$-point functions
of the two states are related by 
\begin{equation}
\langle \Phi_\Mb(x_1)\cdots\Phi_\Mb(x_n)\rangle_{\psi^*\omega} =
\langle \Phi_\Nb(\psi(x_1))\cdots\Phi_\Nb(\psi(x_n))\rangle_{\omega}
\label{eq:npointpullback}
\end{equation}
(adopting an `unsmeared' notation). That is,
the $n$-point function of $\psi^*\omega$ is simply the pull-back of the
$n$-point function of $\omega$ by $\psi$ (or more precisely, by the
duplication of $\psi$ across $n$ copies of $\Mb$). This has an important
consequence when the state $\omega$ is Hadamard, i.e.,
$\omega\in\Ss_\Nb$: because the Hadamard condition is based on the
local metric and causal structure, both of which are preserved by
$\psi$, it is clear that $\psi^*\omega$ is also Hadamard. (A
more elegant proof of this~\cite{BrFrVe03} is to use Radzikowski's characterisation of
the Hadamard condition in terms of the wave-front set of the two-point
function~\cite{Radz96}, and the transformation properties of the
wave-front set under pull-backs.)
This may be expressed by the inclusion $\psi^*\Ss_\Nb\subset \Ss_\Mb$. 

As noted above, the expectation values of the stress-energy tensor is
also constructed in a purely local fashion from the two-point function
of the state. It therefore follows that
\begin{equation}
\langle T_{\Mb\,ab}(x)\rangle_{\psi^*\omega} = \frac{\partial y^{a'}}{\partial
x^a}\frac{\partial y^{b'}}{\partial x^b}\langle T_{\Nb\,a'b'}(y)\rangle_{\omega} 
\end{equation}
where we have written $y=\psi(x)$: like the $n$-point functions, the
expected stress-energy tensor in
state $\psi^*\omega$ is simply the pull-back of that in state $\omega$. In coordinate-free
notation we may write
\begin{equation}
\langle \Ts_\Mb\rangle_{\psi^*\omega} =\psi^*\langle \Ts_\Nb\rangle_{\omega}\,.
\label{eq:Tpullback}
\end{equation}
In the above equations we have written the stress-energy tensor as if it
is an element of the algebra $\Af_\Mb$, which it is not. One may
proceed in two ways: either interpreting Eq.~\eqref{eq:Tpullback} as
the extension of Eq.~\eqref{eq:exp_under_pullback} to an algebra of
Wick polynomials which contains $\Af_\Mb$ as a subalgebra, and in
which $\Ts_{\Mb}$ may be defined as a locally covariant
field~\cite{Ho&Ru02,Ho&Wa01}. For our purposes, however, it
will be simpler to define the smeared stress-energy tensor only
through its expectation values; more technically, we think of it as a 
linear functional on the space of Hadamard states, with the notation
$\langle \Ts_\Mb(\fs)\rangle_\omega$ expressing the value of this
functional applied to the state $\omega$. This has the advantage that
one may deal with all Hadamard states, rather than those which extend
to the Wick algebra~\cite{Ho&Ru02}. 

We emphasise the fact that states are pulled back in this setting;
although one could push forward a state $\omega\in\Ss_\Mb$ to obtain a state
on $\Af_{\psi(\Mb)}$, there is no guarantee that this can be extended to 
a Hadamard state on $\Af_\Nb$, and indeed, such extensions do not
always exist. For example, the Rindler vacuum state on the Rindler wedge is Hadamard in the
interior of the wedge~\cite{Sa&Ve00}, 
but cannot be extended to a Hadamard state on the whole of
Minkowski because its stress-energy tensor diverges at the boundary of
the wedge. See~\cite{FeKaVe05} for further discussion of these issues.

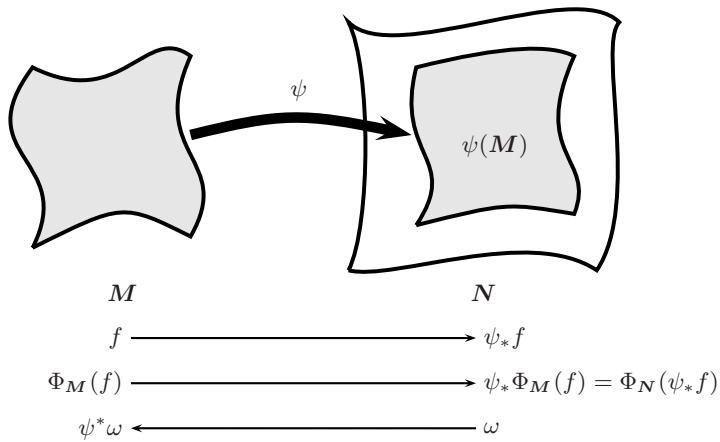
\begin{figure}[h]
\begin{center}
\psset{xunit=3mm,yunit=3mm,runit=3mm}
\begin{pspicture}(-15,-5)(15,15)

\newgray{palegray}{.9}

\pscustom[linewidth=1.5pt,fillstyle=solid,fillcolor=palegray]{
   \psbezier{-}(-12,4)(-10,8)(-15,8)(-12,12)
   \psbezier{-}(-12,12)(-8,10)(-7,14)(-5,12.5)
   \psbezier{-}(-5,12.5)(-7,8)(-3,8)(-5,4.5)
   \psbezier{-}(-5,4.5)(-8,6)(-8,6)(-12,4)
}
\rput*(-8,2){$\bm M$}

\pscustom[linewidth=1.5pt]{
   \psbezier{-}(2,3)(3,5)(3,12)(2,14)
   \psbezier{-}(2,14)(4,12)(12,16)(14,14)
   \psbezier{-}(14,14)(13,13)(15,6)(13,3)
   \psbezier{-}(13,3)(10,5)(4,2.5)(2,3)
}
\rput*(8,2){$\bm N$}

\pscustom[linewidth=1.5pt,fillstyle=solid,fillcolor=palegray]{
   \psbezier{-}(5,5)(7,8)(4,8)(5,12)
   \psbezier{-}(5,12)(9,13)(11,13)(12,12.5)
   \psbezier{-}(12,12.5)(10,8)(13,8)(12,5.5)
   \psbezier{-}(12,5.5)(9,6)(9,6)(5,5)
}
\rput(8.5,8.5){$\psi({\bm M})$}

\psbezier[linewidth=4pt]{->}(-5,9)(-1,10)(0,10)(4.8,9)
\rput*(-0.2,11){$\psi$}

\rput[r](-8,0){\rnode{f}{$f$}}
\rput[l](8,0){\rnode{pf}{$\psi_* f$}}
\ncline[nodesep=3pt]{->}{f}{pf}

\rput[r](-8,-2){\rnode{phi}{$\Phi_{\bm M}(f)$}}
\rput[l](8,-2){\rnode{pphi}{$\psi_*\Phi_{\bm M}(f)=\Phi_{\bm N}(\psi_* f)$}}
\ncline[nodesep=3pt]{->}{phi}{pphi}

\rput[r](-8,-4){\rnode{pw}{$\psi^*\omega$}}
\rput[l](8,-4){\rnode{w}{$\omega$}}
\ncline[nodesep=3pt]{->}{w}{pw}

\end{pspicture} 
\end{center}
\caption{If $\Mb$ is embedded in $\Nb$ by a causal isometry $\psi$, then test
functions and smeared fields may be pushed forwards from $\Mb$ to
$\Nb$, while states (and expectation values) are pulled back from
$\Nb$ to $\Mb$.}  
\end{figure}

\subsection{QEIs in a locally covariant setting}\label{sec:QEI_loc_cov}

We now introduce two types of locally covariant QEIs. A more abstract
(and general) definition can be given in the language of
categories---this will be pursued elsewhere. Recall that a set of
sampling tensors on a globally hyperbolic spacetime is a set of
compactly supported distributions on smooth second rank covariant
tensor fields.

\begin{Def} A {\em locally covariant absolute QEI} assigns to each globally
hyperbolic spacetime $\Mb$ a set of sampling tensors $\FF_\Mb$ on $\Mb$
and a map $\QQ_\Mb:\FF_\Mb\to\RR$ such that (i) we have
\begin{equation}
\langle \Ts_\Mb(\fs)\rangle_\omega \ge -\QQ_\Mb(\fs)
\label{eq:absQIform} 
\end{equation}
for all $\fs\in\FF_\Mb$ and $\omega\in\Ss_\Mb$, and (ii) if
$\psi:\Mb\to\Nb$ is a causal isometric embedding then $\psi_*\FF_\Mb\subset
\FF_\Nb$ and 
\begin{equation}
\QQ_\Mb(\fs) = \QQ_\Nb(\psi_*\fs)
\end{equation}
for all $\fs\in\FF_\Mb$. (We might also express this in the form
$\QQ_\Mb=\psi^*\QQ_\Nb$.)

A {\em locally covariant difference QEI} assigns to each globally
hyperbolic $\Mb$ a set of sampling tensors $\FF_\Mb$ as before, and a map
$\QQ_\Mb:\FF_\Mb\times\Ss_\Mb\to\RR$ such that (i)
\begin{equation}
\langle \Ts_\Mb(\fs)\rangle_\omega -\langle \Ts_\Mb(\fs)\rangle_{\omega_0} \ge -\QQ_\Mb(\fs,\omega_0)
\end{equation}
for each $\fs\in\FF_\Mb$ and all $\omega,\omega_0\in\Ss_\Mb$; (ii)
\begin{equation}
\QQ_\Mb(\fs,\psi^*\omega_0) = \QQ_\Nb(\psi_*\fs,\omega_0)
\label{eq:diffQtransf}
\end{equation}
holds for all $\fs\in\FF_\Mb$ and $\omega_0\in\Ss_\Nb$. 
\end{Def}

We will shortly give examples of each type: Flanagan's two-dimensional
QEIs for massless fields~\cite{Flan02} will be exhibited as a locally covariant
absolute QEI, while (generalisations of) the QEI obtained
in~\cite{Fews00} provide examples of locally covariant difference QEIs.
Before that, let us examine some simple consequences of these
definitions. 

First, suppose that $\Mb'$ is a c.e.g.h.s.\ of $\Mb$, so the identity
map $\imath:\Mb'\to\Mb$ is a causal isometric embedding, and we must have
$\imath_*\FF_{\Mb'}\subset\FF_\Mb$ and
$\QQ_{\Mb'}(\fs)=\QQ_{\Mb}(\imath_*\fs)$. It is sensible to drop the
identity mappings, and write the above in the form
\begin{equation}
\FF_{\Mb'}\subset\FF_\Mb\,,\quad{\rm and}\quad
\QQ_{\Mb'}(\fs)=\QQ_{\Mb}(\fs)~\hbox{for all $\fs\in\FF_{\Mb'}$.}
\end{equation}
If $\psi:\Mb'\to\Nb$ is a causal isometric embedding we then obtain
\begin{equation}
\QQ_{\Mb}(\fs)=\QQ_{\Mb'}(\fs)=\QQ_{\Nb}(\psi_*\fs)
\end{equation}
for all $\fs\in\FF_{\Mb'}$. As one would expect, this shows that
locally covariant absolute QEIs are indifferent to the larger spacetime;
one obtains the same bound whether one is in $\Mb'$ or its image $\Nb'$
in $\Nb$. Although this barely extends the original definition, it is
worth isolating it as a separate result.

\begin{Prop} Suppose a c.e.g.h.s.\ $\Mb'$ of $\Mb$ is causally
isometric to a c.e.g.h.s.\ $\Nb'$ of $\Nb$ under the map $\psi$. Then
a locally covariant absolute QEI obeys
\begin{equation}
\QQ_\Mb(\fs) = \QQ_\Nb(\psi_*\fs)
\end{equation}
for all $\fs\in\FF_{\Mb'}\subset\FF_\Mb$. 
\end{Prop}

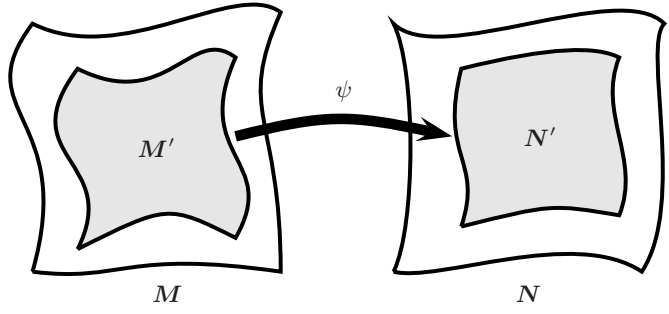
\begin{figure}
\begin{center}
\psset{xunit=3mm,yunit=3mm,runit=3mm}
\begin{pspicture}(-15,1)(15,15)

\newgray{palegray}{.9}

\pscustom[linewidth=1.5pt]{
   \psbezier{-}(-14,3)(-12,10)(-17,10)(-14,14)
   \psbezier{-}(-14,14)(-10,12)(-5,16)(-3,14.5)
   \psbezier{-}(-3,14.5)(-5,10)(-3,10)(-3,3)
   \psbezier{-}(-3,3)(-10,4)(-10,2.5)(-14,3)
}
\rput*(-8,2){$\bm M$}

\pscustom[linewidth=1.5pt,fillstyle=solid,fillcolor=palegray]{
   \psbezier{-}(-12,4)(-10,8)(-15,8)(-12,12)
   \psbezier{-}(-12,12)(-8,10)(-7,14)(-5,12.5)
   \psbezier{-}(-5,12.5)(-7,8)(-3,8)(-5,4.5)
   \psbezier{-}(-5,4.5)(-8,6)(-8,6)(-12,4)
}
\rput(-8.5,8.5){${\bm M}'$}

\pscustom[linewidth=1.5pt]{
   \psbezier{-}(2,3)(3,5)(3,12)(2,14)
   \psbezier{-}(2,14)(4,12)(12,16)(14,14)
   \psbezier{-}(14,14)(13,13)(15,6)(13,3)
   \psbezier{-}(13,3)(10,5)(4,2.5)(2,3)
}
\rput*(8,2){$\bm N$}

\pscustom[linewidth=1.5pt,fillstyle=solid,fillcolor=palegray]{
   \psbezier{-}(5,5)(6,8)(4,8)(5,12)
   \psbezier{-}(5,12)(9,13)(11,13)(12,12.5)
   \psbezier{-}(12,12.5)(11,10)(13,8)(12,5.5)
   \psbezier{-}(12,5.5)(9,6)(9,6)(5,5)
}
\rput(8.5,9){${\bm N}'$}

\psbezier[linewidth=4pt]{->}(-5,9)(-1,10)(0,10)(4.6,9)
\rput*(-0.2,11){$\psi$}

\end{pspicture} 
\end{center}
\caption{Diagram showing the various spacetimes and embeddings in Sec.~\ref{sec:QEI_loc_cov}}
\end{figure}

Let us now examine locally covariant difference QEIs in this situation. Given arbitrary
Hadamard states $\omega_\Mb\in\Ss_\Mb$ and $\omega_\Nb\in\Ss_\Nb$ on the
parent spacetimes $\Mb$ and $\Nb$, there are states $\imath^*\omega_\Mb$
and $\psi^*\omega_\Nb$ in $\Ss_{\Mb'}$, i.e., Hadamard states on
$\Af_{\Mb'}$. Applying the difference QEI to $\psi^*\omega_\Nb$ with $\imath^*\omega_\Mb$
as reference state, we find
\begin{equation}
\langle \Ts_{\Mb'}(\fs)\rangle_{\psi^*\omega_\Nb} 
-\langle \Ts_{\Mb'}(\fs)\rangle_{\imath^*\omega_\Mb} \ge
-\QQ_{\Mb'}(\fs,\imath^*\omega_\Mb)
=-\QQ_{\Mb}(\imath_*\fs,\omega_\Mb)
\end{equation}
where we have used the transformation property Eq.~\eqref{eq:diffQtransf}.
On the other hand, we could equally well apply the difference QEI to
$\imath^*\omega_\Mb$, with $\psi^*\omega_\Nb$
as reference state, to obtain
\begin{equation}
\langle \Ts_{\Mb'}(\fs)\rangle_{\imath^*\omega_\Mb}-
\langle \Ts_{\Mb'}(\fs)\rangle_{\psi^*\omega_\Nb} 
\ge -\QQ_{\Mb'}(\fs,\psi^*\omega_\Nb)
=-\QQ_{\Nb}(\psi_*\fs,\omega_\Nb)
\end{equation}
Combining these inequalities yields
\begin{equation}
\QQ_{\Nb}(\psi_*\fs,\omega_\Nb)\ge \langle \Ts_{\Mb'}(\fs)\rangle_{\psi^*\omega_\Nb} 
-\langle \Ts_{\Mb'}(\fs)\rangle_{\imath^*\omega_\Mb} \ge -\QQ_{\Mb}(\imath_*\fs,\omega_\Mb)\,;
\end{equation}
we may also use the covariance of $\Ts$ to reexpress the central member of
this inequality in terms of expectation values on $\Nb$ and $\Mb$, rather
than $\Mb'$. The result, on dropping identity mappings from the notation, is the following.

\begin{Prop} Suppose a c.e.g.h.s.\ $\Mb'$ of $\Mb$ is causally
isometric to a c.e.g.h.s.\ $\Nb'$ of $\Nb$ under the map $\psi$. Then
a locally covariant difference QEI obeys
\begin{equation}
\QQ_{\Nb}(\psi_*\fs,\omega_\Nb)\ge \langle \Ts_{\Nb}(\psi_*\fs)\rangle_{\omega_\Nb} 
-\langle \Ts_{\Mb}(\fs)\rangle_{\omega_\Mb} \ge -\QQ_{\Mb}(\fs,\omega_\Mb)\,,
\label{eq:main_bd}
\end{equation}
for all $\fs\in\FF_{\Mb'}\subset\FF_\Mb$ and any $\omega_\Mb\in\Ss_\Mb$, $\omega_\Nb\in\Ss_\Nb$. 
\end{Prop}

Note that the QEIs used are those
associated with the full spacetimes $\Mb$ and $\Nb$; similarly, the
states $\omega_\Mb$, $\omega_\Nb$ are states of the field on the full
spacetimes. However, the isometry $\psi$ connects only portions of the
spacetime together and the restriction on the support of $\fs$ is
therefore crucial: in general the above result will not hold when
sampling extends outside the isometric region. It is also worth noting
that we have both lower and upper bounds.

In this paper, we will study the simplest possible setting for this
result, in which $\Mb$ is Minkowski spacetime and $\omega_\Mb$ is the
Minkowski vacuum state. However other situations are possible. For
example, Marecki~\cite{Marecki05} has employed our framework in the case where $\Mb$ is
the exterior Schwarzschild spacetime and $\omega_\Mb$ is the Boulware vacuum.
In the Minkowski case, the result simplifies
because the renormalised stress-energy tensor vanishes in the state $\omega_\Mb$, 
and we have the following statement.

\begin{Cor}\label{Cor:diffMink}
 Suppose a c.e.g.h.s.\ $\Mb'$ of Minkowski space $\Mb$ is causally
isometric to a c.e.g.h.s.\ $\Nb'$ of $\Nb$ under the map $\psi$. Then
a locally covariant difference QEI obeys
\begin{equation}
\QQ_{\Nb}(\psi_*\fs,\omega_\Nb)\ge \langle \Ts_{\Nb}(\psi_*\fs)\rangle_{\omega_\Nb} 
\ge -\QQ_{\Mb}(\fs,\omega_\Mb)\,,
\label{eq:Mink_bd}
\end{equation}
for all $\fs\in\FF_{\Mb'}\subset\FF_\Mb$ and any 
$\omega_\Nb\in\Ss_\Nb$, where $\omega_\Mb$ is the Minkowski vacuum state.
\end{Cor}

\subsection{A locally covariant absolute QEI for massless fields in two
dimensions}\label{sect:lcaQEIs_examples}

The QEI we now describe was originally developed by
Flanagan~\cite{Flan97} for the massless scalar field in two
dimensional Minkowski space, in work which was subsequently
generalised to curved spacetimes~\cite{Voll00,Flan02,Fews04} and also
to arbitrary unitary positive energy conformal
field theories in two dimensional Minkowski space~\cite{Fe&Ho05}. The
results of~\cite{Flan02} were obtained
for two-dimensional spacetimes globally conformal to the whole of
Minkowski space; as noted in~\cite{Fews04}, however, any point of a
globally hyperbolic two-dimensional spacetime has a (causally embedded) neighbourhood which
is conformal to the whole of Minkowski space, and to which Flanagan's
result applies. 

We first state the result of~\cite{Flan02}, and then show that it
meets our definition of a locally covariant absolute QEI. Let
$\Mb$ be a globally hyperbolic two-dimensional spacetime, and suppose
that $\gamma$ is a smooth, future-directed timelike curve, parametrised by proper time
$\tau\in I$, which is completely contained within a c.e.g.h.s. $\Mb'$ of
$\Mb$, such that $\Mb'$ is
globally conformal to the whole of two-dimensional Minkowski space.
Then all Hadamard states
$\omega$ on $\Mb$ obey the QEI
\begin{equation}
\int_I \langle T_{\Mb\,ab}u^a u^b\rangle_\omega(\gamma(\tau))
g(\tau)^2\,d\tau
\ge -\frac{1}{6\pi}\int_I \left[ g'(\tau)^2 + 
g(\tau)^2\left\{R_{\Mb}(\gamma(\tau))-a^c(\tau)a_c(\tau)\right\}\right]\,d\tau
\label{eq:2dcurvedQI}
\end{equation}
for any smooth, real-valued $g$ compactly supported in $I$~\footnote{The
proof employed in~\cite{Flan02,Flan97} proceeds by defining a
function $V:\supp g\to\RR$ with $V'(v)=1/g(v)^2$, and therefore only
applies in the first instance to the
case where $g$ has connected support and no zeros in the interior
thereof. We extend the result to more general $g\in\CoinX{I;\RR}$ by
choosing a nonnegative $\chi\in\CoinX{I}$ with no zeros in the interior
of its support, assumed to be connected, and which is equal to unity on the support of
$g$. Applying Flanagan's result to $g_{\epsilon}(\tau) =
\sqrt{g(\tau)^2+\epsilon^2\chi(\tau)^2}$, we may take the limit $\epsilon\to
0$ to obtain Eq.~\eqref{eq:2dcurvedQI} (cf.\ Cor.~A.2
in~\cite{Fe&Ho05}, where the notation $G=g^2$ is used).}, where $u^a$ is the two-velocity of
$\gamma$, $a^c$ is its acceleration and $R_\Mb$ is the scalar curvature on
$\Mb$. [Note that~\cite{Flan02} uses conventions in which $u^a u_a<0$ for timelike
$u^a$; the bound is therefore modified slightly.]

As we now describe, Flanagan's bound is a locally covariant absolute QEI.
Given $I$, $\gamma$ and $g$ as above, we may define a compactly
supported distribution
$\fs_{I,\gamma,g}$ acting on smooth
second rank covariant tensor fields $\ts$, by
\begin{equation}
\fs_{I,\gamma,g}(\ts) = \int_I g(\tau)^2 u^au^b
t_{ab}|_{\gamma(\tau)}\,d\tau\,.
\label{eq:smearing2d}
\end{equation}
Our set of sampling tensors $\FF^{\rm conf}_\Mb$ (`conf' abbreviating
`conformal') will be the set of all distributions formed in this way. 
[A distribution of the form $\fs_{I,\gamma,g}$ is singularly supported on the curve
$\gamma$; we could also write it in the form
\begin{equation}
f_{I,\gamma,g}^{ab}|_p = \int_I u^au^b g(\tau)^2\delta_{\gamma(\tau)}(p) \,d\tau
\end{equation}
where $\delta_q(p)$ is the $\delta$-function at $q$, obeying $\int_\Mb
\delta_q(p)F(p)\dvol=F(q)$.]

The QEI bound $\QQ^{\rm conf}_\Mb$ is then defined by
\begin{equation}
\QQ^{\rm conf}_\Mb(\fs)= 
\frac{1}{6\pi}\int_I \left[ g'(\tau)^2 +
g(\tau)^2\left\{R_{\Mb}(\gamma(\tau))-a^c(\tau)a_c(\tau)\right\}\right]\,d\tau\
\label{eq:Qdef_2d}
\end{equation}
for any $I$, $\gamma$, $g$ for which $\fs=\fs_{I,\gamma,g}$.
For this to make sense, we must ensure that 
the right-hand side is unchanged if we replace $I$, $\gamma$
and $g$ by $\widetilde{I}$, $\widetilde{\gamma}$ and $\widetilde{g}$
such that
$\fs_{I,\gamma,g}=\fs_{\widetilde{I},\widetilde{\gamma},\widetilde{g}}$.
Since $\gamma$ and $\gamma'$ are both assumed to be parametrised by
proper time, our two sampling tensors must be related in a simple way:
$\supp \widetilde{g}$ is the translation of $\supp g$ by some $\tau_0$,
so that $\widetilde{\gamma}(\tau)=\gamma(\tau+\tau_0)$,
$\widetilde{g}(\tau)^2=g(\tau+\tau_0)^2$ for all
$\tau\in\supp\widetilde{g}$. The only possible ambiguity stems from the
fact that $\widetilde{g}(\tau)$ and $g(\tau+\tau_0)$ mught differ by a
relative sign which can change at zeros of $\widetilde{g}$ of infinite
order. However, it is simple to show that, nevertheless,
$\widetilde{g}'(\tau)^2=g'(\tau+\tau_0)^2$~\footnote{We have
$\widetilde{g}(\tau)^2=g(\tau+\tau_0)^2$, and want to show that
$\widetilde{g}'(\tau)^2=g'(\tau+\tau_0)^2$ for all $\tau$. Differentiating and squaring
yields
$4\widetilde{g}(\tau)^2\widetilde{g}'(\tau)^2=4g(\tau+\tau_0)^2g'(\tau+\tau_0)^2$
from which it follows that $\widetilde{g}'(\tau)^2=g'(\tau+\tau_0)^2$
except perhaps at zeros of $\widetilde{g}(\tau)$. If $\widetilde{g}$
vanishes in a neighbourhood of $\tau$ then so must $\widetilde{g}'$ and
the result holds trivially. For the remaining case, choose a sequence
$\tau_n\to\tau$ with $\widetilde{g}(\tau_n)\not=0$; since
$\widetilde{g}'(\tau_n)^2=g'(\tau_n+\tau_0)^2$, we conclude the required
result by continuity as $n\to\infty$.}, ensuring that the right-hand
side of Eq.~\eqref{eq:Qdef_2d} is unchanged under the reparametrization
of $\fs$.   

The bound Eq.~\eqref{eq:2dcurvedQI} now takes the form of
Eq.~\eqref{eq:absQIform}, so it remains only to verify that $\FF^{\rm conf}_\Mb$ and $\QQ^{\rm conf}_\Mb$ have the
required transformation properties. Suppose $\psi:\Mb\to\Nb$ is a causal
isometric embedding. The push-forward $\psi_*$ acts on $\fs_{I,\gamma,g}\in\FF^{\rm conf}_\Mb$
so that, for any smooth tensor field $t_{ab}$ on $\Nb$, we have
\begin{eqnarray}
(\psi_*\fs_{I,\gamma,g})(\ts) &=& \fs_{I,\gamma,g}(\psi^*\ts)  \nonumber\\
&=& \int_I g(\tau)^2 u^au^b(\psi^*\ts)_{ab}|_{\gamma(\tau)} \,d\tau\nonumber\\
&=& \int_I g(\tau)^2 (\psi^*\us)^a(\psi^*\us)^b t_{ab}|_{\gamma(\tau)}
\,d\tau\nonumber\\
&=& \fs_{I,\psi\circ\gamma,g}(\ts)\,.
\end{eqnarray}
Now the image curve $\psi\circ\gamma$ can certainly be enclosed in a
c.e.g.h.s.\ of $\Nb$ which is conformal to the whole of Minkowski space:
namely the image under $\psi$ of that which
enclosed $\gamma$. Moreover, the image curve has velocity $\psi^*\us$.
It is therefore clear that $\psi_*\fs_{I,\gamma,g}=\fs_{I,\psi\circ\gamma,g}$ is a legitimate sampling tensor
in $\FF^{\rm conf}_\Nb$, so we have shown that $\psi_*\FF^{\rm conf}_\Mb\subset \FF^{\rm conf}_\Nb$. 
It is obvious that $\QQ^{\rm conf}_\Mb(\fs)=\QQ^{\rm conf}_\Nb(\psi_*\fs)$ because all
quantities involved in the bound are invariant under the isometry.

We have thus shown that two-dimensional massless fields obey a
locally covariant absolute QEI. One need not restrict to worldline
averages such as those described above: see~\cite{Flan97,Flan02} for
averages along spacelike or null curves, and~\cite{Fe&Ho05} for
worldvolume averages [in Minkowski space]. We summarise as follows:

\begin{Thm} \label{thm:2d_abs_QWEI}
Let $\Mb$ be a two-dimensional globally hyperbolic spacetime and
let $\Ss_\Mb$ be the class of Hadamard states of the massless Klein--Gordon field on $\Mb$.
Let $\FF^{\rm conf}_\Mb$ consist of all
sampling tensors of the form Eq.~\eqref{eq:smearing2d} where
(i)~$\gamma:I\to\Mb$ is a smooth future-directed timelike curve parametrised
by proper time, with velocity $\us=\dot{\gamma}$; (ii)~$\gamma$ which may be enclosed
in a c.e.g.h.s.\ of $\Mb$ globally conformal to the whole of Minkowski
space; (iii)~$g\in\CoinX{I;\RR}$.   
Then, defining $\QQ^{\rm conf}_\Mb(\fs)$ by Eq.~\eqref{eq:Qdef_2d} for any $I$, $\gamma$, $g$
for which $\fs=\fs_{I,\gamma,g}$, the absolute QWEI
\begin{equation}
\int_\gamma 
\langle T_{\Mb\,ab}\rangle_\omega u^au^bg(\tau)^2\,d\tau
\ge -\QQ^{\rm conf}_\Mb(\fs_{I,\gamma,g})
\end{equation}
holds for all $\omega\in\Ss_\Mb$ and $\fs_{I,\gamma,g}\in\FF^{\rm conf}_{\Mb}$, and is locally covariant.
\end{Thm}

\subsection{Examples of locally covariant difference QEIs}\label{sect:lcdQEIs_examples}

We now give two related examples of locally covariant difference QEIs,
based on methods first introduced in~\cite{Fews00}. The first is a
quantum null energy inequality (QNEI), constraining averages of
the null-contracted stress-energy tensor along timelike
curves~\cite{Fe&Ro03}, while the second is a quantum weak energy
inequality (QWEI), constraining averages of the energy density along timelike
curves~\cite{Fews00}. 

Suppose that $\Mb$ is any globally hyperbolic spacetime of dimension
$d\ge 2$, and $\gamma:I\to M$ is any smooth, future-directed timelike curve. Suppose
further that $k^a$ is a smooth nonzero null vector field defined near $\gamma$. 
Then for any smooth, real-valued $g$, compactly supported in $I$, there
is a difference QNEI~\cite{Fe&Ro03},
\begin{equation}
\int_\gamma 
\left[\langle T_{\Mb\,ab}\rangle_\omega- 
\langle T_{\Mb\,ab}\rangle_{\omega_0}\right]k^ak^bg(\tau)^2\,d\tau
\ge -\int_0^{\infty}\frac{d\alpha}{\pi} \widehat{F}_{\gamma,g,\ks,\omega_0}(-\alpha,\alpha) 
\label{eq:QNEI}
\end{equation}
for all $\omega,\omega_0\in\Ss_\Mb$, where the hat denotes Fourier
transform and
\begin{equation}
F_{\gamma,g,\ks,\omega_0}(\tau,\tau') = g(\tau)g(\tau') 
\langle \nabla_{\ks}\Phi_\Mb(\gamma(\tau))\nabla_{\ks}\Phi_\Mb(\gamma(\tau'))\rangle_{\omega_0}\,.
\end{equation}
in which we have written $\nabla_\ks$ for $k^a\nabla_a$. 
[More precisely, the last factor is a distributional pull-back of the differentiated
two-point function. We also adopt the nonstandard convention
\begin{equation}
\widehat{f}(\lambda) = \int dt \, e^{i\lambda t} f(t)
\end{equation}
for Fourier tranforms; for purposes of comparison, we note that
the same convention was used in~\cite{Fews00}, but not in~\cite{Fe&Ro03}.]
The integral on the right-hand side of Eq.~\eqref{eq:QNEI} is finite as
a consequence of $\omega_0$ being Hadamard. We emphasise that there is
no necessity for $\omega$ and $\omega_0$ to be represented as vectors or
density matrices in a common
Hilbert space representation in order to prove the QEIs described in
this section, because the proof may be phrased entirely in the algebraic
formulation of QFT. 

The above result was derived in~\cite{Fe&Ro03} based on an earlier result
in~\cite{Fews00}, described below. However, it is slightly easier
to show that it is locally covariant, which is why we have presented it
first. To accomplish our task, we define $\FF^{\rm null}_\Mb$ to consist of 
all compactly supported distributions $\fs_{I,\gamma,\ks,g}$ on smooth 
second rank covariant tensor fields $\ts$ on $M$,
such that
\begin{equation}
\fs_{I,\gamma,\ks,g}(\ts) = \int_I g(\tau)^2 k^ak^b t_{ab}\,d\tau
\label{eq:smearing}
\end{equation}
for $\gamma$, $k^a$, $g$ obeying the conditions already mentioned in
this subsection {\em and} with $g$ having connected support with no
zeros of infinite order in its interior, for reasons to be explained
shortly. We write $\tCoinX{I;\RR}$ for the set of functions $g$ of this type. 
As in the two-dimensional case 
it is clear that the assignment $\Mb\to\FF^{\rm null}_\Mb$ is covariant in
the required sense. 

The QEI bound is then defined by setting $\QQ^{\rm null}_\Mb(\fs,\omega_0)$ equal
to minus the right-hand side of Eq.~\eqref{eq:QNEI}, for any $I$,
$\gamma$, $\ks$ and
$g$ such that $\fs=\fs_{I,\gamma,\ks,g}$. The particular parametrisation is
not important, for reasons similar to those explained in the previous
subsection. However here it is important that $g\in\tCoinX{I;\RR}$:
otherwise we could change $g$ to $h(\tau)=\sigma(\tau)g(\tau)$ with
$\sigma$ changing sign from $+1$ to $-1$ at a zero of $g$ of infinite
order, say at $\tau_0$; although $\fs_{I,\gamma,\ks,h}=\fs_{I,\gamma,\ks,g}$, the two
functions $F_{\gamma,h,\ks,\omega_0}$ and $F_{\gamma,g,\ks,\omega_0}$
differ when, for example, $\tau<\tau_0<\tau'$. (The restriction to
$\tCoinX{I;\RR}$ is not, however, very significant because it is dense
in $\CoinX{I;\RR}$, as is shown in Appendix~\ref{appx:smooth}.)
Finally,  
the covariance property Eq.~\eqref{eq:diffQtransf} 
follows because Eq.~\eqref{eq:npointpullback} (for the case $n=2$) implies 
\begin{equation}
F_{\gamma,g,\ks,\psi^*\omega_0}(\tau,\tau') = 
F_{\psi\circ\gamma,g,\psi_*\ks,\omega_0}(\tau,\tau')\,.
\end{equation}
We summarise what has been proved.
\begin{Thm}\label{Thm:4d_QNEI}
Let $\Mb$ be a globally hyperbolic spacetime of dimension $d\ge 2$ and
let $\Ss_\Mb$ be the class of Hadamard states of the Klein--Gordon field
of mass $m\ge 0$ on $\Mb$. Let $\FF^{\rm null}_\Mb$ consist of all
sampling tensors of the form Eq.~\eqref{eq:smearing} where
$\gamma:I\to\Mb$ is a smooth future-directed timelike curve parametrised
by proper time, $\ks$ is a smooth nonzero null field defined near the
track of $\gamma$ and $g\in\tCoinX{I;\RR}$. For each $\fs\in\FF^{\rm
null}_\Mb$ and reference state $\omega_0\in\Ss_\Mb$ define
\begin{equation}
\QQ^{\rm null}_\Mb(\fs,\omega_0)
=\int_0^{\infty}\frac{d\alpha}{\pi} \widehat{F}_{\gamma,g,\ks,\omega_0}(-\alpha,\alpha) 
\end{equation}
for any $I$, $\gamma$, $\ks$, $g$ with $\fs=\fs_{I,\gamma,\ks,g}$. 
Then the difference QNEI
\begin{equation}
\int_\gamma 
\left[\langle T_{\Mb\,ab}\rangle_\omega- 
\langle T_{\Mb\,ab}\rangle_{\omega_0}\right]k^ak^bg(\tau)^2\,d\tau
\ge -\QQ^{\rm null}_\Mb(\fs,\omega_0)
\end{equation}
holds for all $\omega,\omega_0\in\Ss_\Mb$ and $\fs_{I,\gamma,\ks,g}\in\FF^{\rm
null}_{\Mb}$, and is locally covariant.
\end{Thm}

Our second example of a locally covariant difference QEI constrains the
energy density. We keep $\gamma$ and $g$ as before, but replace $k^a$ by
the velocity $u^a$ of the trajectory. Then the following difference QWEI
holds for all $\omega,\omega_0\in\Ss_\Mb$~\cite{Fews00}:
\begin{equation}
\int_\gamma 
\left[\langle T_{\Mb\,ab}\rangle_\omega- 
\langle T_{\Mb\,ab}\rangle_{\omega_0}\right]u^au^bg(\tau)^2\,d\tau
\ge -\int_0^{\infty}\frac{d\alpha}{\pi} \widehat{G}_{\gamma,g,\eb,\omega_0}(-\alpha,\alpha) 
\label{eq:QWEI}
\end{equation}
where 
\begin{equation}
G_{\gamma,g,\eb,\omega_0}(\tau,\tau') =\frac{1}{2} 
g(\tau)g(\tau')\left[ \delta^{\mu\mu'}
\langle \nabla_{\es_\mu}\Phi_\Mb(\gamma(\tau))\nabla_{\es_{\mu'}}\Phi_\Mb(\gamma(\tau'))\rangle_{\omega_0}
+m^2\langle \Phi_\Mb(\gamma(\tau))\Phi_\Mb(\gamma(\tau'))\rangle_{\omega_0}\right]
\end{equation}
and $\eb=(e_\mu^a)_{\mu=0\ldots,d-1}$ is a smooth $d$-bein defined in a
neighbourhood of $\gamma$ with $e_0^a=u^a$ on $\gamma$. 

The frame $\eb$ adds a new ingredient to the discussion of covariance,
which was not explored in~\cite{Fews00}. 
Subject to the condition $e_0^a|_{\gamma}=u^a$, 
any choice of $\eb$ will give a QEI bound, which may have differing numerical values. 
When considering a causal isometry $\psi:\Mb\to\Nb$, we must therefore
find a way of choosing frames in the two spacetimes so as to give equal
values to the QWEI bound, in accordance with covariance. 
One solution would be to incorporate the frame as part of the data in
the QWEI, [i.e.,
writing $\QQ^{\rm weak}_\Mb(\fs,\eb,\omega_0)$ and using the
push-forward $\psi_*\eb$ on $\psi(\Mb)$] but this seems rather inelegant. 
Fortunately, a better solution is at hand: it turns out that we can
covariantly specify a subclass of frames guaranteed to yield the same
numerical bound. This is accomplished by requiring, in addition to
$e_0^a|_{\gamma}=u^a$, that the
$d$-bein $\eb$ be invariant under Fermi--Walker transport along $\gamma$,
i.e.,
\begin{equation}
\frac{D_{\rm FW} e_\mu^a}{d\tau} \equiv u^b\nabla_b e_\mu^a + a_b e_\mu^bu^a- u_b e_\mu^ba^a = 0
\end{equation}
for each $\mu=0,\ldots,d-1$, where $a^a$ is the acceleration of
$\gamma$. If $\eb'$ is another $d$-bein also invariant under
Fermi--Walker transport and with ${e'}_0^a=e_0^a=u^a$, then it must be
that $\eb'$ is related to $\eb$ by a rigid rotation along $\gamma$,
i.e., ${e'}_i^a|_{\gamma(\tau)}=S_{i}^{\phantom{i}j}e_j^a|_{\gamma(\tau)}$ for some fixed $S\in{\rm
SO}(d-1)$, because Fermi--Walker transport preserves inner products.
It is now easy to see that
$G_{\gamma,g,\eb,\omega_0}=G_{\gamma,g,\eb',\omega_0}$, because the form
of $\eb$ {\em off} the curve $\gamma$ is irrelevant, provided it is smooth. Accordingly this
QEI depends only on the smearing tensor $\fs$ [defined by analogy with
Eq.~\eqref{eq:smearing}] and the reference state. 

We emphasise that this is only one method of constructing a locally
covariant bound in this setting, and others may be convenient in other
contexts. For example, it would be possible to simply take the infimum
of the bound over all $d$-beins with $\es_0=\us$; this is certainly
locally covariant, but impractical for calculational purposes. 

With this detail addressed, it is now straightforward to show that this
QEI is locally covariant by exactly the same arguments as used in the
null-contracted case, and the additional observation that $\psi_*\eb$ is
Fermi--Walker transported along $\psi\circ\gamma$ if $\eb$ is along
$\gamma$. Again, we summarise what has been established.

\begin{Thm} \label{thm:4d_QWEI}
Let $\Mb$ be a globally hyperbolic spacetime of dimension $d\ge 2$ and
let $\Ss_\Mb$ be the class of Hadamard states of the Klein--Gordon field
of mass $m\ge 0$ on $\Mb$. Let $\FF^{\rm weak}_\Mb$ consist of all
sampling tensors of the form Eq.~\eqref{eq:smearing2d} where
$\gamma:I\to\Mb$ is a smooth future-directed timelike curve parametrised
by proper time and with velocity $\us=\dot{\gamma}$, and $g\in\tCoinX{I;\RR}$.
For each $\fs\in\FF^{\rm weak}_\Mb$ and reference state $\omega_0\in\Ss_\Mb$ define
\begin{equation}
\QQ^{\rm weak}_\Mb(\fs,\omega_0)
=\int_0^{\infty}\frac{d\alpha}{\pi} \widehat{G}_{\gamma,g,\eb,\omega_0}(-\alpha,\alpha)
\label{eq:dQWEIbound}
\end{equation}
for any $I$, $\gamma$, $g$ with $\fs=\fs_{I,\gamma,g}$,
and any smooth tetrad $\eb$ defined near the track of $\gamma$ with $\eb_0|_\gamma=\us$ 
and which is invariant under Fermi--Walker transport along $\gamma$.  
Then the difference QWEI
\begin{equation}
\int_\gamma 
\left[\langle T_{\Mb\,ab}\rangle_\omega- 
\langle T_{\Mb\,ab}\rangle_{\omega_0}\right]u^au^bg(\tau)^2\,d\tau
\ge -\QQ^{\rm weak}_\Mb(\fs,\omega_0)
\end{equation}
hold for all $\omega,\omega_0\in\Ss_\Mb$ and $\fs_{I,\gamma,g}\in\FF^{\rm
weak}_{\Mb}$, and is locally covariant.
\end{Thm}

Most cases considered in the sequel will actually involve averages in
static spacetimes along timelike curves which are static
trajectories (i.e., orbits of a hypersurface orthogonal timelike Killing field $\xis$) and with
$\omega_0$ chosen to be a static Hadamard state (with respect to the
same Killing field). In these cases the bounds derived above simplify considerably, because
the two-point function of $\omega_0$ obeys
\begin{equation}
w_2(\psi_t x,\psi_t x') = w_2(x,x')
\end{equation}
for any $t$, $x,x'$, where $\psi_t$ is the one-parameter group of
isometries obtained from $\xis$. We fix a particular orbit
$\gamma(\tau)=\psi_\tau(x_0)$, which may be assumed to be a proper-time
parametrisation [as $\xi_a\xi^a$ is constant along $\gamma$ and may be
set equal to unity]. Then the two-point function, restricted to
$\gamma$, can be expressed as
\begin{equation}
w_2(\gamma(\tau),\gamma(\tau')) = w_2(\psi_\tau (x_0), \psi_{\tau'}(x_0))=
w(\tau-\tau')
\end{equation}
where $w(\tau)=w_2(\psi_\tau(x_0),x_0)$. The same
time-translational invariance is obtained for derivatives
$(\nabla_\vs\otimes\nabla_\vs)w_2)(\gamma(\tau),\gamma(\tau'))$,
provided that $\vs$ is invariant under the Killing flow, or
equivalently, has vanishing Lie derivative with respect to $\xis$ on
$\gamma$, i.e.,
$\pounds_\xis\vs|_\gamma=0$. 

This simplifies the QWEI bound~\eqref{eq:QWEI} as follows. If $\eb$ is
Lie-transported along $\gamma$ then
\begin{equation}
G_{\gamma,g,\eb,\omega_0}(\tau,\tau')=g(\tau)g(\tau')T_{\gamma,\omega_0}(\tau-\tau')
\label{eq:G_in_taudiff}
\end{equation}
holds for some `single variable' distribution
$T_{\gamma,\omega_0}$; moreover, $\eb$ is also invariant under
Fermi--Walker transport along $\gamma$ (owing to hypersurface orthogonality of $\xis$
\footnote{Let $f=\xi^a\xi_a$, and note that the curve $\gamma$ has
acceleration $a^a=-\frac{1}{2}\nabla^a f$. Suppose $\pounds_\xis\vs=0$. Then $\xi^a\nabla_a
v^b=v^a\nabla_a \xi^b$, which permits us to write the Fermi--Walker derivative
as $(D_{\rm FW}\vs/dt)_b = v^a(\nabla_a\xi_b+\xi_{[a}\nabla_{b]}\log f)$. But
this vanishes for hypersurface orthogonal $\xis$; see, e.g., Appendix
C.3 in~\cite{Wald_GR}.}.) Then, as shown in~\cite{Fews00,Fe&V03}, the
QEI Eq.~\eqref{eq:QWEI} becomes
\begin{equation}
\int_\gamma 
\left[\langle T_{\Mb\,ab}\rangle_\omega- 
\langle T_{\Mb\,ab}\rangle_{\omega_0}\right]u^au^bg(\tau)^2\,d\tau
\ge -\int_{-\infty}^{\infty} du\, \left| \widehat{g}(u) \right|^2 Q_{\gamma,\omega_0}(u)\,,
\label{eq:QWEI_withQ}
\end{equation}
where $Q_{\gamma,\omega_0}(u)$ is a positive polynomially bounded function defined by
\begin{equation}
Q_{\gamma,\omega_0}(u) = \frac{1}{2\pi^2} \int_{(-\infty,u)} dv\,
\widehat{T}_{\gamma,\omega_0}(v). 
\label{eq:Q_function}
\end{equation}
Additionally, if $\omega_0$ is a ground state (as was the case
in~\cite{Fews00}) one may show that
$\widehat{T}_{\gamma,\omega_0}(\sigma)=0$ for $\sigma<0$, and so
the function $Q_{\gamma,\omega_0}$ is supported on the positive
half-line only. More generally, it is always the case that 
$\widehat{T}_{\gamma,\omega_0}(\sigma)$ decays rapidly as
$\sigma\to-\infty$, so $Q_{\gamma,\omega_0}$ is always well-defined~\cite{Fe&V03}.
Technically, $\widehat{T}_{\gamma,\omega_0}(\sigma)$ is a measure, and
may have $\delta$-function spikes which would exhibit themselves as
discontinuities in $Q_{\gamma,\omega_0}(u)$. Since we define
$Q_{\gamma,\omega_0}(u)$ as an integral over the open interval
$(-\infty,u)$, it is continuous from the left.

A similar analysis holds for the QNEI Eq.~\eqref{eq:QNEI}, provided that
the null vector field $\ks$ has vanishing Lie derivative along $\gamma$,
$\pounds_\xi \ks=0$, because we have 
\begin{equation}
F_{\gamma,g,\ks,\omega_0}(\tau,\tau')=g(\tau)g(\tau')S_{\gamma,\ks,\omega_0}(\tau-\tau')
\end{equation}
for some distribution $S_{\gamma,\ks,\omega_0}$.

To conclude this section, we mention that more general QEI bounds may be constructed along similar lines,
based on other decompositions of the contracted stress--energy tensor as
a sum of squares. This includes bounds averaged over spacetime volumes,
see, e.g.~\cite{Fews05}. However we will not need this generality here, and
observe only that one would need to ensure that such decompositions are
made in a canonical fashion to obtain a locally covariant bound. 

\section{Applications: General examples}\label{sec:general_apps}

In this section we develop some simple consequences of the QEIs
described in Secs.~\ref{sect:lcaQEIs_examples}
and~\ref{sect:lcdQEIs_examples}, specialised to Minkowski
space. These will then be utilised in more general spacetimes using the
local covariance properties of these bounds. Our results are obtained by
converting QEI bounds into eigenvalue problems which can then be solved. 

For the most part, we will consider the scalar field of mass $m\ge 0$ on $d$-dimensional
globally hyperbolic spacetimes for $d\ge 2$; special features of
massless fields in two dimensions will be treated in Sec.~\ref{sect:GEmassless2d}.
Accordingly, let $\Nb$ be a $d$-dimensional globally hyperbolic
spacetime, and let $\Mb_d$ denote $d$-dimensional
Minkowski space. As illustrated in
Fig.~\ref{fig:Mink_embed}, let $\gamma:I\to\Nb$ be a smooth,
future-directed timelike curve, parametrised by proper time $\tau\in I$,
and assume $\gamma$ may be enclosed in a c.e.g.h.s.\ $\Nb'$ of $\Nb$ so that $\Nb'$
is the image of a c.e.g.h.s.\ $\Mb'$ of $\Mb_d$ under a causal
isometric embedding $\psi:\Mb'\to\Nb$. Thus the curve $\gamma$ is the image
of a curve $\widetilde{\gamma}(\tau) = \psi^{-1}(\gamma(\tau))$ in
$\Mb_d$; because $\psi$ is an isometry, $\tau\mapsto\widetilde{\gamma}(\tau)$ is also a proper time
parametrisation, and $\widetilde{\gamma}$ has the same proper
acceleration as $\gamma$ for each $\tau\in I$. 

Given any $g\in\tCoinX{I;\RR}$, define a sampling tensor on Minkowski
space $\fs\in\FF^{\rm weak}_{\Mb_d}$ by 
\begin{equation}
\fs(\ts)=\int_I t_{ab}|_{\widetilde{\gamma}(\tau)} \widetilde{u}^a \widetilde{u}^b g(\tau)^2\,d\tau
\label{eq:fdef}
\end{equation}
on smooth covariant rank-two tensor fields $\ts$ on $\Mb_d$, where
$\widetilde{\us}$ is the velocity of $\widetilde{\gamma}$. [Recall that
$g\in\tCoinX{I;\RR}$ means that $g$ is a real-valued smooth function
whose support is compact, connected and contained in $I$, and that $g$
has no zeros of infinite order in the interior of its support.] Under the
isometry, $\fs$ is mapped to $\psi_*\fs$, with action
\begin{equation}
\psi_*\fs(\ts)=\int_I t_{ab}|_{\gamma(\tau)} u^a u^b g(\tau)^2\,d\tau\,,
\end{equation}
where $\ts$ is now any smooth covariant rank-$2$ tensor field on $\Nb$.
Applied to the stress-energy tensor, $\psi_*\fs$ therefore provides a weighted average of
the energy density along $\gamma$. 
Our aim is to place constraints on these averages using the locally covariant difference QWEI given in
Theorem~\ref{thm:4d_QWEI}. By local covariance, Cor.~\ref{Cor:diffMink} guarantees
that 
\begin{equation}
\int_I \langle T_{\Nb\,ab}\rangle_\omega(\gamma(\tau))u^a u^b
g(\tau)^2\,d\tau = \langle \Ts_{\Nb}(\psi_*\fs)\rangle_\omega
\ge -\QQ^{\rm weak}_{\Mb_d}(\fs,\omega_{\Mb_d})\,,
\label{eq:loccovimp}
\end{equation}
where $\omega_{\Mb_d}$ is the Minkowski vacuum state. 

We will be particularly interested in the least upper bound of the energy density along $\gamma$,
\begin{equation}
\EE := \sup_\gamma \langle T_{\Nb\,ab} u^a u^b\rangle_\omega\,.
\end{equation}
Since the energy density is
smooth, this value must be the maximum value taken by the field on the
closure of the track of $\gamma$. Using the trivial estimate $\EE\ge \langle T_{\Nb\,ab} u^a
u^b\rangle_\omega(\gamma(\tau))$ for each $\tau\in I$, we have
\begin{equation}
\EE \int_I g(\tau)^2\,d\tau \ge \int_I \langle T_{\Nb\,ab}\rangle_\omega(\gamma(\tau))u^a u^b
g(\tau)^2\,d\tau
\end{equation}
and, putting this together with Eq.~\eqref{eq:loccovimp}, we obtain the inequality 
\begin{equation}
\EE \int_I g(\tau)^2\,d\tau \ge  -\QQ^{\rm
weak}_{\Mb_d}(\fs,\omega_{\Mb_d})\,,
\label{eq:fundamental_E_bound}
\end{equation}
which holds, in the first place, for all $g\in\tCoinX{I;\RR}$. In the
next two subsections we will analyse this in two special cases: namely,
inertial motion and uniform acceleration.

\subsection{Inertial curves}\label{sect:GA_inertial}

When $\gamma$ is inertial the QWEI of Theorem~\ref{thm:4d_QWEI} takes
the simpler form described in Eqs.~\eqref{eq:QWEI_withQ} and~\eqref{eq:Q_function}
above~\cite{Fe&E98}:
\begin{equation}
\QQ^{\rm weak}_{\Mb_d}(\fs,\omega_{\Mb_d}) = K_d
\int_{m}^\infty \frac{du}{\pi}\, u^d |\widehat{g}(u)|^2  Q_d(u/m)\,,
\label{eq:QWEI_d_massive}
\end{equation}
where 
\begin{equation}
Q_d(x) =\frac{d}{x^d}\int_1^x dy\,  y^2(y^2-1)^{(d-3)/2}\,,
\end{equation}
and the constant $K_d$ is $K_d=A_{d-2}/(2d(2\pi)^{d-1})$, where $A_k$
is the area of the unit $k$-sphere. (Notation varies slightly from that
used in~\cite{Fe&E98}.)

For all $d\ge 3$, it is clear that $Q_d(x)\le 1$ for all $x\ge 1$, while
one may show that $Q_2(x)<1.2$ on the same domain~\footnote{To see this, we note that
$Q_2(\cosh\alpha) = \tanh\alpha + \alpha(1-\tanh^2\alpha)$
and that the maximum of this expression on $\RR^+$ occurs at the
(unique) solution to $\alpha\tanh\alpha=1$, which is
numerically $\alpha_0 = 1.199679$. Now $Q_2(\cosh\alpha_0)=\alpha_0$, so
we find that $Q_2(x)\le\alpha_0<1.2$ on $[1,\infty)$ as claimed.}. 
Using these results, we may estimate Eq.~\eqref{eq:QWEI_d_massive}
rather crudely by
\begin{equation}
\QQ^{\rm weak}_{\Mb_d}(\fs,\omega_{\Mb_d}) \le K_d'
\int_{0}^\infty \frac{du}{\pi}\, u^d |\widehat{g}(u)|^2 
\label{eq:crude_massive_bound}
\end{equation}
with $K_d'=K_d$ for $d\ge 3$ and $K_2'=1.2K_2$. Note that we have made
two changes here: (a) $Q_d(u/m)$ has been replaced by unity; (b) the
lower integration limit $m$ has been replaced by zero. 

We now specialise to
even dimensions $d=2k$, $k\ge 1$. Because $g$ is
real-valued, $|\widehat{g}(u)|$ is even and we may write 
\begin{equation}
\int_{0}^\infty \frac{du}{\pi}\, u^d |\widehat{g}(u)|^2
=\int_{-\infty}^\infty \frac{du}{\pi}\, u^{2k} |\widehat{g}(u)|^2
=\int_I d\tau |(D^kg)(\tau)|^2\,,
\end{equation}
where $D$ is the differential operator $D=-id/d\tau$ and we have used
Parseval's theorem, and the fact that $g$ vanishes outside $I$. 

Inserting the above in Eq.~\eqref{eq:fundamental_E_bound}, we have shown that $\EE$ obeys the inequality
\begin{equation}
\EE \int_I |g(\tau)|^2\,d\tau \ge -K_d' 
\int_I d\tau |(D^kg)(\tau)|^2
\label{eq:E_bound}
\end{equation}
for all $g\in\tCoinX{I;\RR}$. The class $\tCoinX{I;\RR}$ is inconvenient to work
with directly; fortunately, the same
inequality holds for general $g\in\CoinX{I}$, as we now show. First, any
$g\in\CoinX{I;\RR}$ is the limit of a sequence of $g_n\in\tCoinX{I;\RR}$
for which $g_n\to g$ and $D^kg_n\to D^kg$ in $L^2(I)$ (see
Appendix~\ref{appx:smooth}). Applying the above inequality to each
$g_n$, we may take the limit $n\to\infty$ to conclude that it holds for
$g$ as well. Having established the result for arbitrary real-valued
$g\in\CoinX{I}$, we extend to general complex-valued $g$ by applying it
to real and imaginary parts separately, and then adding. Accordingly the
inequality Eq.~\eqref{eq:E_bound} holds for all $g\in\CoinX{I}$.

Integrating by parts $k$ times, and noting that no boundary terms arise
because $g$ vanishes near the boundary $\partial I$ of $I$,
Eq.~\eqref{eq:E_bound} may be rearranged to give
\begin{equation}
-\frac{\EE}{K_d'} \le \frac{\ip{g}{L g}}{\ip{g}{g}}\,,
\label{eq:Ritz}
\end{equation}
where $\ip{\cdot}{\cdot}$ denotes the usual $L^2$-inner product on
$I$, and the operator $L=(-1)^k d^{2k}/d\tau^{2k}$ on $\CoinX{I}$. 
Our aim is now to minimise the right-hand side over the class of
$g$ at our disposal (excluding the identically zero function). Now the
operator $L$ is symmetric~\footnote{An operator $A$ is symmetric on a
domain $\DD$ if $\ip{\psi}{A\varphi}=\ip{A\psi}{\varphi}$
for all $\psi,\varphi\in\DD$, which shows that the adjoint $A^*$ agrees
with $A$ on $\DD$, but does not exclude the possibility that $A^*$ has a
strictly larger domain of definition than $A$.} and positive, i.e.,
$\ip{g}{Lg}\ge 0$ for all $g\in\CoinX{I}$. By Theorem~X.23
in~\cite{RSvII:75}), the solution to our minimisation problem is the
lowest element $\lambda_0$ of the spectrum of $\widehat{L}$, the
so-called {\em Friedrichs extension} of $L$. This is a self-adjoint
operator with the same action as $L$ on $\CoinX{I}$, but which is
defined on a larger domain in $L^2(I)$. In particular, every function in
the domain of $\widehat{L}$ obeys the boundary condition $g=g'=\cdots=g^{(k-1)}=0$ at $\partial
I$. (See~\cite{Fe&Te00}, where the technique of reformulating quantum
energy inequalities as eigenvalue problems was first introduced, and
which contains a self-contained exposition of the necessary
operator theory.) One
may think of this as a precise version of the Rayleigh--Ritz principle.
Once we have determined $\lambda_0$, we then have the bound
\begin{equation}
\EE \ge -\lambda_0 K_d'\,,
\end{equation}
so the problem of determining the lower bound is reduced to
the analysis of a Schr\"odinger-like equation, subject to the boundary
conditions mentioned above. 

The two examples of greatest interest to us are $k=1$ and $k=2$,
representing two- and four-dimensional spacetimes. Starting with $k=1$, 
let us suppose that
$I$ is the interval $(-\tau_0/2,\tau_0/2)$ for some $\tau_0>0$. We
therefore solve $-g''=\lambda g$ subject to Dirichlet boundary
conditions at $\pm\tau_0/2$; as is well known, the lowest eigenvalue is 
$\lambda_0=\pi^2/\tau_0^2$ and corresponds to the eigenfunction $g(\tau)=\cos \pi
\tau/\tau_0$. [A possible point of confusion is that, if $g$ is
extended so as to vanish outside $I$, it will not be smooth. However
there is no contradiction here: the point is that the infimum is not
attained on $\CoinX{I}$.] Thus we have
\begin{equation}
\EE \ge -\frac{3\pi}{10\tau_0^2}\,,
\label{eq:emax2di}
\end{equation}
because $K_2'=1.2 K_2=1.2/(4\pi)=3/(10\pi)$ [by convention, the zero-sphere has
area $A_0=2$]. We may infer, without further calculation, that the bound must be
zero if $I=\RR$, because (returning to the Ritz quotient
Eq.~\eqref{eq:Ritz}), the infimum over all
functions in $\CoinX{\RR}$ must be less than or equal to the infimum
over all functions in $\CoinX{I}$ for any bounded $I$ (a similar
argument applies to the semi-infinite case). Thus $\lambda_0$ can be no
greater than zero; on the other hand, the minimum cannot be negative
either, because the original functional is nonnegative. Accordingly Eq.~\eqref{eq:emax2di} holds in
all cases, with $\tau_0$ equal to the length of the interval $I$.

In the four-dimensional case $k=2$, we proceed in a similar way, solving 
$g''''=\lambda g$ subject to $g=g'=0$ at $\partial I$. In the case where $I$ is bounded,
$I=(-\tau_0/2,\tau_0/2)$ [without loss of generality], the spectrum
consists only of positive eigenvalues. It is easy to see that 
the solutions to the eigenvalue equation $g''''=\beta^4 g$ are linear
combinations of trigonometric and
hyperbolic functions. The lowest eigenfunction solution which obeys
the boundary conditions is
\begin{equation}
g(\tau) = \cosh(\beta \tau/\tau_0)  - \frac{\cosh(\beta /2)}{\cos(\beta /2)}
\cos(\beta \tau/\tau_0)
\end{equation}
where $\beta \approx 4.730040745$ is the minimum positive solution to
\begin{equation}
\tan(\beta/2) = -\tanh(\beta/2).    
\label{eq:beta_def}
\end{equation}
Since $K_4'=1/(16\pi^2)$, we obtain
\begin{equation}
\EE \ge -\frac{500.5639}{16 \pi^2 \tau_0^4}=
-\frac{3.169858}{\tau_0^4}
\end{equation}
If $I$ is semi-infinite or infinite, we may argue exactly as in the
two-dimensional case that the bound vanishes, in agreement with the
formal limit $\tau_0\to\infty$.

Clearly this approach will give similar results in any even dimension,
with a consequent increase in complexity in solving the eigenvalue
problem. Nonetheless, it is clear that the resulting bound will always
scale as $\tau_0^{-d}$. In fact, this is even true in odd spacetime dimensions,
where the eigenvalue problem would involve a nonlocal operator and is
not easily tractable.

We summarise what has been proved so far in the following way.
\begin{Prop} \label{Prop:emaxmassive}
Let $\Nb$ be a globally hyperbolic 
spacetime of dimension $d\ge 2$ and suppose that a timelike geodesic segment $\gamma$ of
proper duration $\tau_0$ may be enclosed in a c.e.g.h.s.\ of $\Nb$
which is causally isometric to a c.e.g.h.s.\ of Minkowski space $\Mb_d$, then
\begin{equation}
\sup_\gamma \langle T_{\Nb\,ab} u^a u^b\rangle_\omega \ge -\frac{C_d}{\tau_0^d}\,,
\label{eq:emaxmassive}
\end{equation}
for all Hadamard states $\omega$ of the Klein--Gordon field of mass
$m\ge 0$ on $\Nb$. The constants $C_d$ depend only on $d$. 
In particular, $C_2=3\pi/10=0.942478\ldots$, while $C_4=3.169858\ldots$.
\end{Prop}

{\noindent\bf Remark:} When the field has nonzero mass, we can expect rather more rapid
decay than given by this estimate.
To see why, return to the argument leading to
Eq.~\eqref{eq:crude_massive_bound}. If we reinstate $m$ as the lower integration
limit, we have
\begin{equation}
\QQ^{\rm weak}_{\Mb_d}(\fs,\omega_{\Mb_d}) \le K_d'
\int_{m}^\infty \frac{du}{\pi}\, u^d |\widehat{g}(u)|^2 
\label{eq:better_massive_bound}
\end{equation}
Suppose for simplicity that $I=(-\tau_0/2,\tau_0/2)$. 
If we write $g_{\tau_0}(\tau) = \tau_0^{-1/2}g_0(\tau/\tau_0)$, for
$g_0\in\CoinX{-1/2,1/2}$, a change of variables yields
\begin{equation}
\QQ_{\Mb_d}(\fs, \omega_{\Mb_2})\le
\frac{K_d' G_d(m\tau_0)}{\tau_0^d}
\end{equation}
where the nonnegative quantity
\begin{equation}
G_d(x) =\int_x^\infty \frac{dy}{\pi} y^d|\widehat{g}_0(y)|^2
\end{equation}
decays rapidly as $x\to\infty$, owing to the rapid decay of
$\widehat{g}$. Thus the estimate Eq.~\eqref{eq:emaxmassive}
is quite crude when $m\tau_0\gg 1$; it is hoped to return to this
elsewhere. 

Equipped with Prop.~\ref{Prop:emaxmassive}, we may now address the first two
examples presented in the Introduction. First, the proposition asserts
that no Hadamard state can maintain
an energy density lower than $-C_d/\tau_0^d$ for proper time $\tau_0$
along an inertial curve in a Minkowskian c.e.g.h.s.\ of $\Nb$. In
particular, this justifies the claim made in Example 1
in the Introduction. 

Our bounds clearly depend only on $\tau_0$, which in turn is controlled
by the size of the Minkowskian region $\Nb'$. By choosing the curve
$\gamma$ and $\Nb'$ in an appropriate way, fairly simple geometrical considerations
can thus provide good {\em a priori}
bounds on the magnitude and duration of negative energy density. A good
illustration is the following (which includes Example 2 in the Introduction). 

Suppose that a $d$-dimensional globally hyperbolic spacetime $\Nb$ 
with metric $\gb$ is stationary with respect to timelike
Killing vector $t^a$ and admits the smooth foliation into constant time surfaces
$N\cong\RR\times\Sigma$. Suppose there is a (maximal) subset $\Sigma_0$ of
$\Sigma$, with nonempty interior, for which $\gb$ takes the Minkowski
form on $\RR\times \Sigma_0$. Choose any point $(t,x)$ in $N$, with
$x\in\Sigma_0$ and suppose that we may isometrically embed a Euclidean
$(d-1)$-ball of radius $r$ in $\Sigma_0$, centred at $x$ (see
Fig.~\ref{fig:example2}). Then the interior
of the double cone $J^+(\{(t-r,x)\})\cap J^-(\{(t+r,x)\})$ is a c.e.g.h.s. of $\Nb$
which is isometric to a c.e.g.h.s. of Minkowski space, and contains an inertial
curve sgement $\gamma(\tau)=(\tau,x)$ parametrised by the interval $(t-r,t+r)$
of proper time. Any Hadamard state $\omega$ on $\Nb$ therefore obeys
\begin{equation}
\sup_\gamma \langle T_{\Nb\,ab} u^a u^b\rangle_\omega 
\ge -\frac{C_d}{(2r)^d}
\end{equation}
along $\gamma$. Writing $r(x)$ for the minimum distance from $x$ to the
boundary of $\Sigma_0$, it is clear that this inequality holds for all
$r<r(x)$ and hence, by continuity, for $r=r(x)$. Moreover, 
if the state is stationary [for example,
if it is the ground state], then the energy density takes 
a constant value along $\gamma$ and we obtain
\begin{equation}
\langle T_{ab}n^a n^b\rangle_{\omega}(t,x)\ge -\frac{C_d}{(2r(x))^d}
\end{equation}
for any $x\in\Sigma_0$, where $n^a$ is the unit vector along $t^a$.
In this way we obtain a universal bound on the fall-off of negative
energy densities in such spacetimes, which could be used to provide a
quantitative check on exact calculations, if these are possible, or to
provide some precise information in situations where they are not. The
bound is of course very weak close to the boundary of $\Sigma_0$: this does not
imply that the energy density diverges as this boundary is approached, of
course, but merely indicates that it would not be incompatible with the
quantum inequalities for there to exist geometries on $\RR\times
(\Sigma\backslash \Sigma_0)$ for which
the stationary energy density just outside might be very negative.

To conclude this subsection, let us briefly discuss 
the null-contracted QEI Eq.~\eqref{eq:QNEI} in the present context. For
simplicity, we restrict ourselves to four dimensions. Suppose $\widetilde{k}^a$ is
a nonzero null vector field which is covariantly constant along $\widetilde{\gamma}$, so, in
particular, $\widetilde{u}^a\widetilde{k}_a$ is also constant on $\widetilde{\gamma}$.
Our sampling tensor is now defined to be $\fs\in\FF^{\rm null}_{\Mb_4}$
with action
\begin{equation}
\fs(\ts)=\int_I t_{ab}|_{\widetilde{\gamma}(\tau)} \widetilde{k}^a \widetilde{k}^b g(\tau)^2\,d\tau\,.
\end{equation}
on smooth covariant rank-$2$ tensor fields $\ts$ on $\Mb_4$. In exactly
the same way as for the QWEI discussed above, we may apply local
covariance to the QNEI of Thm.~\ref{Thm:4d_QNEI}, so yielding
\begin{equation}
\int \langle T_{\Nb\,ab}\rangle_\omega k^a k^b g(\tau)^2\,d\tau 
\ge -\QQ_{\Mb_4}^{\rm null}(\fs,\omega_{\Mb_4})\,,
\end{equation}
where, as shown in~\cite{Fe&Ro03},
\begin{equation}
\QQ_{\Mb_4}^{\rm null}(\fs,\omega_{\Mb_4})=
-\frac{(u^ak_a)^2}{12\pi^2}\int_{-\infty}^\infty g''(\tau)^2\,d\tau
\end{equation}
for the massless scalar field (and in fact this bound also constrains
the massive field too). This differs from the corresponding QWEI
by a factor of $4(u^ak_a)^2/3$ [recall that $K_4'=1/(16\pi^2)$], so we may immediately deduce the following result. 
\begin{Prop} \label{prop:nullmax4d}
Let $\Nb$ be a four-dimensional globally hyperbolic
spacetime and suppose that a timelike geodesic segment $\gamma$ of proper duration
$\tau_0$ may be enclosed in a c.e.g.h.s.\ of $\Nb$ which is causally
isometric to a c.e.g.h.s. of Minkowski space. If $k^a$ is a covariantly
constant null vector field on $\gamma$ then we have
\begin{equation}
\sup_{\gamma} \langle T_{\Nb\,ab}\rangle_\omega k^a k^b \ge -\frac{C'_4(u^ak_a)^2}{\tau_0^4}
\end{equation}
for any Hadamard state $\omega$ of the Klein--Gordon field, where $C'_4=4C_4
/3=4.226477\ldots$.
\end{Prop}
This result justifies the claim made above Eq.~(38) of~\cite{Fe&Ro05}, where an
application is presented.

\subsection{Uniformly accelerated trajectories in four dimensions}
\label{sec:unif_accn}

We now turn to the case where $\gamma$ has uniform
constant proper acceleration $\alpha$. For simplicity we consider only massless
fields in four dimensions, but expect similar results in more general
cases. We need to estimate $\QQ^{\rm weak}_{\Mb_4}(\fs,\omega_{\Mb_4})$
where $\fs$ is supported on the uniformly accelerated worldline
$\widetilde{\gamma}$ in $\Mb_4$. It will be convenient to drop the
tilde from $\widetilde{\gamma}$ and the subscript from $\Mb_4$. Without loss of generality, we may
assume $\gamma:I\to\Mb$ is parametrized so that
\begin{equation}
\gamma(\tau)=\left(\begin{array}{c}
\xi_o \sinh(\tau/\xi_o)\\ \xi_o \cosh(\tau/\xi_o)\\ y_o\\ z_o
\end{array}\right)
\qquad\mbox{with}\qquad
u^a(\tau)= \frac{d \gamma(\tau)}{d\tau}^a = \left(\begin{array}{c}
\cosh(\tau/\xi_o)\\ \sinh(\tau/\xi_o)\\0\\ 0\end{array}\right),
\end{equation}
where $\xi_o=\alpha^{-1}$. 

The first step in our calculation is to set up an orthonormal
tetrad field surrounding the worldline,
\begin{equation}
e_0^a = \frac{1}{\sqrt{x^2-t^2}}\left(\begin{array}{c} x\\ t\\ 0\\ 0
\end{array}\right)
,\qquad
e_1^a = \frac{1}{\sqrt{x^2-t^2}}\left(\begin{array}{c} t\\ x\\ 0\\ 0
\end{array}\right)
,\qquad
e_2^a = \left(\begin{array}{c} 0\\ 0\\ 1\\ 0\end{array}\right)
,\qquad
e_3^a = \left(\begin{array}{c} 0\\ 0\\ 0\\ 1\end{array}\right),
\end{equation}
which, satisfies the two properties required: namely, that $e_0^a$ agrees
with the velocity $u^a$ on $\gamma$, and that the frame is
invariant under Fermi--Walker transport along $\gamma$. 
The required bound is then given by
\begin{equation}
\QQ^{\rm weak}_\Mb(\fs,\omega_\Mb)=
\int_0^{\infty}\frac{d\alpha}{2\pi} \widehat{G}_{\gamma,g,\eb,\omega_\Mb}(-\alpha,\alpha) 
\end{equation}
where 
\begin{equation}
G_{\gamma,g,\eb,\omega_\Mb}(\tau,\tau') =\frac{1}{2} 
g(\tau)g(\tau')\delta^{\mu\mu'}
\langle \nabla_{\es_\mu}\Phi_\Mb(\gamma(\tau))\nabla_{\es_{\mu'}}\Phi_\Mb(\gamma(\tau'))\rangle_{\omega_\Mb}\,.
\end{equation}
We evaluate this quantity in stages, beginning by noting that 
\begin{eqnarray}
\lefteqn{\delta^{\mu\mu'}
\langle
\nabla_{\es_\mu}\Phi_\Mb(x)\nabla_{\es_{\mu'}}\Phi_\Mb(x')\rangle_{\omega_\Mb}}\nonumber\\
&=& \left[ \frac{(x x'+ t t')
(\partial_t \partial_{t'}+\partial_x \partial_{x'})+(x t'+ t x')
(\partial_t \partial_{x'}+\partial_x \partial_{t'})}{\sqrt{x^2-t^2}
\sqrt{{x'}^2-{t'}^2}} +\partial_y \partial_{y'}+\partial_z 
\partial_{z'}\right] W_\Mb^{(2)}(x,x')\,,
\label{eq:rindler_Edensity}
\end{eqnarray}
where
\begin{equation}
W_\Mb^{(2)}(x,x')= -\lim_{\epsilon\to 0^+}\frac{1}{4\pi^2}\left[ (t-t'-i\epsilon)^2-
|\xb-\xb'|^2\right]^{-1}
\end{equation}
is the Wightman function of the vacuum state.
Performing the necessary derivatives and pulling back to the
worldline, we obtain, after some calculation,
\begin{equation}
G_{\gamma,g,\eb,\omega_0}(\tau,\tau')=g(\tau)g(\tau')T(\tau-\tau')\,,
\end{equation}
where $T$ is the limit (in the distributional sense)
as $\epsilon\to 0^+$ of
\begin{equation}
T_{\epsilon}(\sigma)=\frac{3}{32\pi^2\xi_o^4}\, \csch^4 \left(
\frac{\sigma}{2\xi_o}-i\frac{\epsilon}{\xi_o} \right).
\end{equation}
Thus we are in the situation of Eq.~\eqref{eq:G_in_taudiff}, and the
bound becomes
\begin{equation}
\QQ^{\rm weak}_\Mb(\fs,\omega_\Mb)=
\int_{-\infty}^{\infty}du\,|\widehat{g}(u)|^2 Q(u)\,,
\label{eq:Q_qfn}
\end{equation}
where 
\begin{equation}
Q(u)=\frac{1}{2\pi^2}\int_{(-\infty,u)} dv\, \widehat{T}(v)\,.
\end{equation}

To obtain the required Fourier transform, we first use contour
integration~\footnote{The contour involved is the rectangle with `long' sides given by the
interval $[-R,R]$ of the real axis, and its translate $[-R,R]+2\pi\xi_o
i$. The contour encloses a single pole, of fourth order, at $z=2i\epsilon$
and the contribution of the `short' sides
vanishes as $R\to\infty$. One also exploits the fact that the
contributions from the two `long' sides are equal up to a factor of
$-e^{-2\pi\xi_o}$.} to find
\begin{equation}
\widehat{T}_\epsilon(u) = \frac{e^{-2u\epsilon}}{2\pi\xi^4_o}\left(\frac{\xi_o^4 u^3 + \xi_o^2 u}
{1-e^{-2\pi\xi_o u}}\right)\,,
\end{equation}
which decays exponentially as $u\to -\infty$, provided
$\epsilon<\pi\xi_o$.
Taking the limit $\epsilon\to 0^+$ it is easy to check that
\begin{equation}
\widehat{T}(u) = \frac{1}{2\pi\xi^4_o}\left(\frac{\xi_o^4 u^3 + \xi_o^2 u}
{1-e^{-2\pi\xi_o u}}\right),
\end{equation}
Note that this Fourier transform has support
on the whole real line, not just the positive half line. Thus
\begin{equation}
Q(u) = \frac{1}{4\pi^3\xi_o^4}\int_{-\infty}^u 
\frac{\xi_o^4 v^3 + \xi_o^2 v}
{1-e^{-2\pi\xi_o v}}\, dv\,.
\end{equation}

Our aim is now to estimate $Q(u)$ in order to obtain a bound which may be
analysed by eigenvalue techniques as in the previous subsection. Beginning in the
half-line $u<0$, we may estimate
\begin{equation}
Q(u)< Q(0) = \frac{11}{960 \pi^3 \xi_o^4}
\label{eq:Qbdneg}
\end{equation}
since $Q(u)$ is everywhere increasing. On the other hand, for $u\ge 0$, we may split
the integral into $Q(0)$ and the contribution from $[0,u]$ to give
\begin{equation}
Q(u) = \frac{11}{960 \pi^3 \xi_o^4} + \frac{1}{4\pi^3\xi_o^4}\left[
\int_0^u (\xi_o^4v^3+\xi_0^2v)\,dv + \int_0^u \frac{\xi_o^4
v^3+\xi_o^2v}{e^{2\pi\xi_o v}-1}\,dv\right]
\end{equation}
after rearranging. Now the last integral is increasing in $u$, so we may bound
it by its limit as $u\to+\infty$, to yield
\begin{equation}
Q(u) \le 
\frac{1}{16\pi^3\xi_o^4}\left(\xi_o^4 u^4+2\xi_o^2 u^2 + \frac{11}{30}\right)
\label{eq:Qbdpos}
\end{equation}
for $u>0$. 
Using the estimates Eqs.~\eqref{eq:Qbdneg} and~\eqref{eq:Qbdpos}, and the fact that
$|\widehat{g}(u)|^2$ is even,
\begin{eqnarray}
\QQ^{\rm weak}_\Mb(\fs,\omega_\Mb) &\le& \frac{11}{960\pi^3\xi_o^4} \int_{-\infty}^0 du\, |\widehat{g}(u)|^2
+\frac{1}{16\pi^3\xi_o^4} \int_{0}^\infty du\, |\widehat{g}(u)|^2 
\left(\xi_o^4 u^4+2\xi_o^2 u^2 + \frac{11}{30}\right)\,.
\nonumber\\
&=& \frac{1}{16\pi^2\xi_o^4}\int_{-\infty}^\infty \frac{du}{2\pi}|\widehat{g}(u)|^2 
\left(\xi_o^4 u^4+2\xi_o^2 u^2 + \frac{11}{20}\right)
\end{eqnarray}
Applying Parseval's theorem, we arrive at
\begin{equation}
\QQ^{\rm weak}_\Mb(\fs,\omega_\Mb) \le \frac{1}{16\pi^2}\int_{-\infty}^\infty
d\tau\,\left(|g''(\tau)|^2+\frac{2}{\xi_o^{2}}|g'(\tau)|^2+\frac{11}{20\xi_o^4}|g(\tau)|^2\right)\,,
\end{equation}
and, together with Eq.~\eqref{eq:fundamental_E_bound}, we now
have
\begin{equation}
\EE \int_I |g(\tau)|^2\,d\tau \ge
\frac{1}{16\pi^2}\int_{-\infty}^\infty
d\tau\,\left(|g''(\tau)|^2+\frac{2}{\xi_o^{2}}|g'(\tau)|^2+\frac{11}{20\xi_o^4}|g(\tau)|^2\right)\,,
\end{equation}
for any $g\in\tCoinX{I;\RR}$. As in the previous subsection, we may
extend this inequality to arbitrary $g\in\CoinX{I}$, and then optimise over
this class. This leads to the conclusion that
\begin{equation}
\EE \ge -\frac{\mu_0}{16\pi^2}\,,
\end{equation}
where $\mu_0$ is the lowest (positive) eigenvalue for the equation
\begin{equation}
g''''-\frac{2}{\xi_o^2}g'' + \frac{11}{20\xi_o^4}g = \mu g
\end{equation}
on $I$, subject to boundary conditions $g=g'=0$ at $\partial I$. 

Let us suppose that $I$ is bounded, writing $I=(-\tau_0/2,\tau_0/2)$
without loss of generality. It is convenient to write
\begin{equation}
\mu =\frac{20\lambda^2-9}{20\xi_o^4}
\end{equation}
for then the eigensolutions must be scalar multiples of
\begin{equation}
g(\tau) =  \cosh\frac{\sqrt{\lambda+1}\tau}{\xi_o} + A\cos\frac{\sqrt{\lambda-1}\tau}{\xi_o}\,,
\end{equation}
where 
\begin{equation}
A = \frac{\sqrt{\lambda+1}
\sinh(\sqrt{\lambda+1}\tau_0/(2\xi_o))}{\sqrt{\lambda-1}
\sin (\sqrt{\lambda-1}\tau_0/(2\xi_o))}
\end{equation}
and $\lambda>1$ solves
\begin{equation}
\sqrt{\lambda+1}\tanh\frac{\sqrt{\lambda+1}\tau_0}{2\xi_o} = -
\sqrt{\lambda-1}\tan\frac{\sqrt{\lambda-1}\tau_0}{2\xi_o}\,.
\label{eq:lambdaeq}
\end{equation}
\begin{figure}
\begin{center}
\scalebox{0.6}{\includegraphics{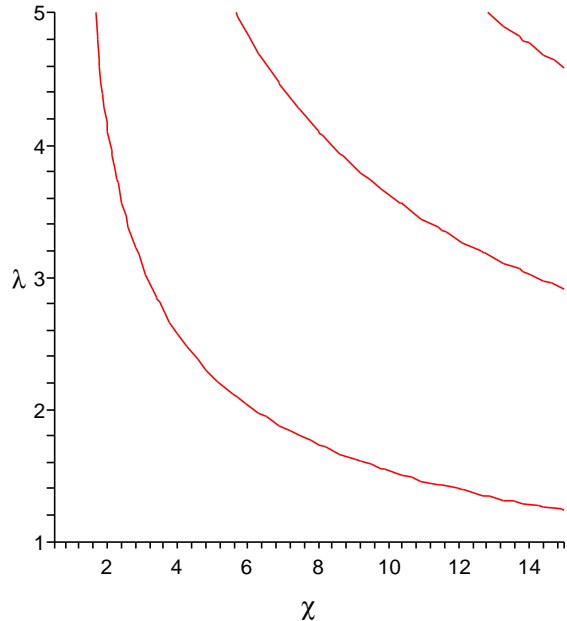}}
\end{center}
\caption{\label{fig:accn_evals} The first three solutions $\lambda$ to
Eq.~\eqref{eq:lambdaeq}, showing the dependence on $\chi=\tau_0/\xi_0$.}
\end{figure}
We shall denote the minimum solution to this equation in $(1,\infty)$ by
$\lambda_0$ (see Fig.~\ref{fig:accn_evals});
clearly $\lambda_0$ depends only on the ratio of the sampling time
$\tau_0$ to the acceleration scale $\xi_o$. Two limits are of
interest. Firstly, when $\tau_0/\xi_0\ll 1$, one may show that
$\lambda_0^2\sim\beta^4\xi_o^4/\tau_0^4$ where $\beta$ is as in Eq.~\eqref{eq:beta_def}. Thus we regain the usual
short-timescale constraint Eq.~\eqref{eq:emaxmassive}. This supports the `usual
assumption' (see~\cite{Fews04} for references) that sampling at scales shorter than those determined by the
acceleration or curvature is governed by the bound obtained for inertial
curves in Minkowski space. On the other hand, if we take $\tau_0\gg
\xi_o$, we see that $\lambda_0\to 1$, so
\begin{equation}
\EE \gtrsim -\frac{11}{320\pi^2\xi_o^4}
\label{eq:crude_acc_AWEC}
\end{equation}
in this limit. 

In fact, more can be said for the inextendible case $\tau_0=\infty$,
because the approximations made to gain Eq.~\eqref{eq:Qbdpos} are
rather wasteful in this limit. 
Choose any $g\in\CoinX{(-1/2,1/2)}$ with $g(0)\not=0$ and define
$g_{\tau_0}(\tau) = \tau_0^{-1/2} g(\tau/\tau_0)$, denoting the corresponding
sampling tensor $\fs_{\tau_0}$. Then a simple change
of variables argument applied to Eq.~\eqref{eq:Q_qfn} shows that
\begin{equation}
\QQ^{\rm weak}_\Mb(\fs_{\tau_0},\omega_\Mb)=\int_{-\infty}^\infty dv\,
|\widehat{g}(v)|^2 Q(v/\tau_0)\,,
\end{equation}
and the limit $\tau_0\to\infty$ may be taken under the
integral sign to yield
\begin{equation}
\lim_{\tau_0\to\infty} \QQ_\Mb(\fs_{\tau_0},\omega_\Mb)
= 2\pi Q(0)\int_{-\infty}^\infty d\tau |g_{\tau_0}(\tau)|^2\,.
\end{equation}
In particular, if $g$ has unit $L^2$-norm, we have
\begin{equation}
\liminf_{\tau_0\to\infty} \frac{1}{\tau_0}\int_\gamma \langle T_{ab}u^a u^b\rangle_\omega
g(\tau/\tau_0)^2\,d\tau
\ge -\frac{11\alpha^4}{480\pi^2}\,,
\label{eq:accelerated_AWEC}
\end{equation}
where $\alpha=\xi_o^{-1}$ is the proper acceleration of the curve, 
as asserted in Example 3 in the Introduction. Thus long term averages of
the energy density measured along the curve are bounded from below, and no
energy density can be less than this bound over the entire worldline.
This is an improvement by a factor of $3/2$ over the bound given in Eq.~\eqref{eq:crude_acc_AWEC}.
Using a more refined analysis one could presumably
extract it as the limit of a result for general $\tau_0$, but we will
not pursue this here. To summarise, we have reached the following
conclusions. 

\begin{Prop} \label{prop:emax_4d_accel}
Let $\Nb$ be a four-dimensional globally hyperbolic
spacetime containing a timelike curve $\gamma$ of proper duration $\tau_0$
and constant proper acceleration $\alpha$. If $\gamma$ may be enclosed
in a c.e.g.h.s.\ of $\Nb$ which is causally isometric to a c.e.g.h.s.\
of Minkowski space then we have
\begin{equation}
\sup_\gamma \langle T_{\Nb\,ab} u^a u^b\rangle_\omega 
\ge -\frac{(20\lambda_0^2-9)\alpha^4}{320\pi^2}
\end{equation}
for any Hadamard state $\omega$ of the massless Klein--Gordon field, where
$\lambda_0$ is the smallest solution to Eq.~\eqref{eq:lambdaeq} in
$[1,\infty)$ and depends on $\alpha\tau_0$. If $\gamma$ has infinite
proper duration, we also have the more stringent constraint Eq.~\eqref{eq:accelerated_AWEC}.
\end{Prop}

\subsection{Massless fields in two dimensions}\label{sect:GEmassless2d}

So far, we have only utilised the locally covariant difference QEIs of
Sec.~\ref{sect:lcdQEIs_examples}. For massless fields in two dimensions,
however, we also have the absolute QEI developed by Flanagan and others,
described in Sec.~\ref{sect:lcaQEIs_examples}, which are also known to be
optimal bounds. In this subsection we briefly
discuss how the results of the previous subsections may be sharpened and
generalised in this context. In fact the formula for the QEI bound is
sufficiently simple that we may work directly in curved spacetime,
rather than in Minkowskian subregions. 

Let $\gamma:I\to \Nb$ be a smooth future-directed timelike curve, with velocity $u^a$ and
accleration $a^c$ in a two-dimensional globally hyperbolic spacetime
$\Nb$. As before, $I$ is an open interval of proper time. In order to
apply Flanagan's bound, we make the additional assumption that $\gamma$
may be enclosed within a c.e.g.h.s.\ $\Nb'$ of $\Nb$, which is globally
conformal to the whole of Minkowski space. Then Flanagan's QEI 
asserts that
\begin{equation}
\int_I \langle T_{\Nb\,ab}u^a u^b\rangle_\omega(\gamma(\tau))
g(\tau)^2\,d\tau
\ge -\frac{1}{6\pi}\int_I \left[ g'(\tau)^2 + g(\tau)^2
\{R_\Nb(\gamma(\tau))-a^c(\tau)a_c(\tau)\} \right]\,d\tau
\end{equation}
for all Hadamard states $\omega$ and any smooth, real-valued $g$
compactly supported in $I$, i.e., $g\in\CoinX{I;\RR}$. 

We proceed as above, obtaining the estimate
\begin{equation}
\EE\int_I g(\tau)^2\,d\tau \ge 
-\frac{1}{6\pi}\int_I \left[ g'(\tau)^2 + 
g(\tau)^2 \{R_\Nb(\gamma(\tau))-a^c(\tau)a_c(\tau)\}\right]\,d\tau\,,
\end{equation}
for all $g\in\CoinX{I;\RR}$, where $\EE = \sup_{\gamma} \langle T_{\Mb\,ab}u^a u^b\rangle_\omega$ as
usual. Converting to an eigenvalue problem, we deduce that 
\begin{equation}
\EE \ge -\frac{\lambda_0}{6\pi}\,,
\end{equation}
where $\lambda_0$ is the lowest element in the spectrum of the
Friedrichs extension of the operator
\begin{equation}
(Lg)(\tau) = -g''(\tau) +(R_\Nb(\gamma(\tau))-a^c(\tau)a_c(\tau))g(\tau)\,,
\end{equation}
on $\CoinX{I}$. Provided $R_\Nb$ and $a^ca_c$ are bounded along
$\gamma$, the correct boundary conditions are Dirichlet conditions $g=0$
on $\partial I$ (see e.g.,~\cite{Fe&Te00}). We now give two illustrative examples.
\begin{Prop} \label{Prop:emax2d'}
Suppose $\Nb$ is a globally hyperbolic two-dimensional
spacetime with a c.e.g.h.s.\ $\Nb'$ which is globally conformal to the
whole of Minkowski space. Then the following hold for all Hadamard states
$\omega$ on $\Nb$:\\
(a) If $\gamma$ is a curve of proper duration $\tau_0$ contained in $\Nb'$,
with $R_{\Nb}-a^ca_c\equiv S$ constant along $\gamma$, then
\begin{equation}
\sup_{\tau\in I} \langle T_{\Nb\,ab}u^a u^b\rangle_\omega(\gamma(\tau))
\ge \frac{S}{6\pi} - \frac{\pi}{6\tau_0^2}\,.
\label{eq:emax2dii'}
\end{equation}
(b) If $\gamma:\RR\to \Nb$ has (signed) proper acceleration growing linearly with
proper time, $d\alpha/d\tau = p$, and $R_\Nb\equiv 0$ on $\gamma$, then 
\begin{equation}
\sup_{\tau\in I} \langle T_{\Nb\,ab}u^a u^b\rangle_\omega(\gamma(\tau)) \ge -\frac{|p|}{6\pi}\,.
\label{eq:emax2diii'}
\end{equation}
\end{Prop}
The proof is straightforward: for~(a), the eigenvalue problem is
$-g''(\tau) = (\lambda+6\pi S) g(\tau)$ on an interval of length $\tau_0$ subject to Dirichlet
boundary conditions, which easily yields the stated
result. For~(c), we may choose the origin of proper time so
that $a^ca_c=-p^2\tau^2$ for some constant $p$. The eigenvalue problem is then
\begin{equation}
-g''(\tau) + p^2\tau^2 g(\tau) = \lambda g(\tau)
\end{equation}
which is the harmonic oscillator equation [and the Friedrichs extension
is also the standard harmonic oscillator Hamiltonian]. The minimum value of
$\lambda$ is therefore the `zero-point' value $\lambda_0=|p|$. [The comparison with the
usual quantum mechanical harmonic oscillator would correspond to units
in which the mass and Planck's constant are both set to $2$.] Thus we
obtain the required result.

\section{Calculations in specific spacetimes}\label{sec:specific_apps}

In this section we illustrate our general method by some concrete
calculations in a variety of locally Minkowskian spacetimes in both two
and four dimensions. For the most part, we focus on the lower bounds,
but upper bound calculations are included where they are
enlightening. For each spacetime we consider, exact values of the
renormalised stress-energy tensor are known (or easily obtained from
existing results) for one or more states. This permits comparison with
the results of our method.

\subsection{Two-dimensional timelike cylinder}\label{sec:2D_cyl}

Consider the massless scalar field on the two-dimensional timelike
cylinder, $\Cb$, i.e., Minkowski
space $\Mb_2$ quotiented by the group of translations $(t,x)\mapsto (t,x+n L)$
($n\in\ZZ$). The Casimir vacuum $\omega_\Cb$ is the ground state of the
scalar field on $\Cb$ (more precisely, it is a state on the algebra of
first derivatives of the field--we will ignore this subtlety, which does
not modify any of our conclusions below). The renormalized
expectation value of the vacuum stress-tensor has the form
\begin{equation}
\langle T_{\Cb\,ab}\rangle_{\omega_\Cb}(x) = \rho_\Cb
\left( \begin{array}{cc}
1 & 0 \\ 0 & 1\\
\end{array}\right),
\label{eq:2d_stresstensor}
\end{equation}
where $\rho_\Cb$ is a constant. Our aim is to use quantum inequalities
to provide upper and lower bounds on $\rho_\Cb$. The value of $\rho_C$
is, of course, well known, and will satisfy the bounds we now derive;
our aim is to demonstrate how it may be bounded without direct calculation.

\begin{figure}
\begin{center}
\includegraphics{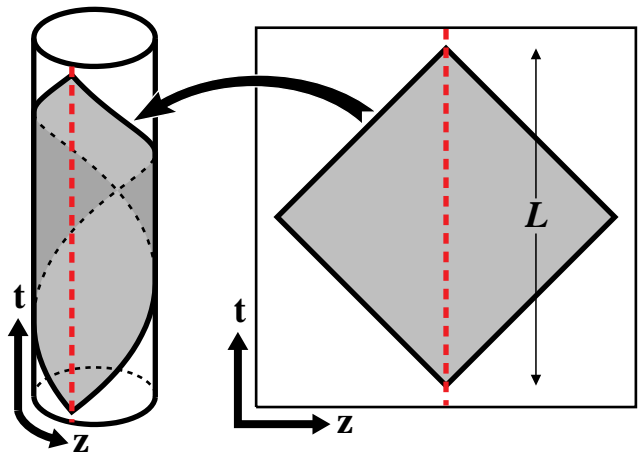} 
\end{center}
\caption{\label{fig:cylinder} Diagram showing the ``largest'' causal
diamond in two dimensional Minkowski space that can be isometrically
embedded into the two dimensional cylinder spacetime with periodicity
$L$ in the $z$-direction.  To an observer inside the diamond in the
cylinder spacetime, the quantum field theory and states would be
indistinguishable from that in Minkowski space.  The dashed vertical
line is the worldline of a stationary observer. }
\end{figure}

In order to apply our method, we must identify suitable globally
hyperbolic subspacetimes of $\Cb$. For any $0<\tau_0\le L$, 
we may define a timelike geodesic $\gamma:(0,\tau_0)\to\Cb$ by
$\gamma(\tau)=(\tau,0)$. Then the 
double cone ${\rm int}\,\left(J^+(\gamma(0))\cap J^-(\gamma(\tau))\right)$
is a
causally embedded globally hyperbolic subspacetime of $\Cb$, containing
$\gamma$. As this subspacetime is globally conformal to the
whole of Minkowski space and the energy density is constant
along $\gamma$, we have the lower bound
\begin{equation}
\rho_\Cb \ge  -\frac{\pi}{6\tau_0^2}\,,
\end{equation}
from Prop.~\ref{Prop:emax2d'}(a) (in the case $S=0$). This bound clearly becomes more stringent as $\tau_0$ is
increased, so we obtain the best bound possible (within this method) by
taking $\tau_0=L$. As shown in Fig.~\ref{fig:cylinder}, the
corresponding diamond is one for which the corners of the diamond just barely
fail to touch on the back of the cylinder.  This gives the final result
\begin{equation}
\rho_\Cb \ge  -\frac{\pi}{6L^2}\,.
\end{equation}


We now demonstrate how to find an upper bound on $\rho_\Cb$ for which we must
employ our locally covariant difference QWEI. Let $\widetilde{\gamma}$
be the curve $\widetilde{\gamma}(\tau)=(\tau,0)$ in $\Mb_2$ and let
$\Mb'={\rm int}\, J^+(\widetilde{\gamma}(0))\cap
J^-(\widetilde{\gamma}(\tau_0))$, for some $0<\tau_0<L$, which is a
c.e.g.h.s.\ of $\Mb_2$. Then the quotient map $q:\Mb_2\to\Cb$ defines a
causal isometric embedding of $\Mb'$ in $\Cb$, with $q(\Mb')$ equal to
the double cone constructed earlier in this subsection. By
Cor.~\ref{Cor:diffMink} we have
\begin{equation}
\langle T_{\Cb}(q_*\fs)\rangle_{\omega_\Cb} \le \QQ^{\rm weak}_\Cb(\fs,\omega_\Cb)
\end{equation}
for any sampling tensor $\fs\in\FF_{\Mb'}^{\rm weak}$. We define $\fs$
by Eq.~\eqref{eq:fdef} for $g\in\tCoinX{(0,\tau_0);\RR}$ and then use
the constancy of the energy density along $\gamma$ to find
\begin{equation}
\rho_\Cb \int_{0}^{\tau_0} g(\tau)^2\,d\tau
\le\QQ^{\rm weak}_\Cb(\fs,\omega_\Cb)\le\frac{1}{2\pi} \int  (g'(\tau))^2 d\tau.
\end{equation}
where the last inequality is derived in Appendix~\ref{sec:QI_cyl}.
As usual, this may be converted into an eigenvalue problem: here,
$\rho_\Cb\le \lambda_0/(2\pi)$ where $\lambda_0=(\pi/\tau_0)^2$ is the minimum
eigenvalue of $-d^2/d\tau^2$ on $(0,\tau_0)$ subject to Dirichlet
boundary conditions. Combining with our earlier lower
bound, we thus have
\begin{equation}
-\frac{\pi}{6 L^2} \le \rho_\Cb \le \frac{\pi}{2 L^2}.
\end{equation}

The known value of $\rho_\Cb$ is exactly $-\pi/(6L^2)$, see \cite{Brl&Dv},
which, remarkably, saturates the lower bound.  Thus we have shown that, 
in the cylinder spacetime, the
Casimir vacuum energy density is the lowest possible static energy
density compatible with the quantum energy inequalities.  This however is not
always the case, which we will see in later examples.

\begin{figure}
\begin{center}
\scalebox{0.6}{\includegraphics{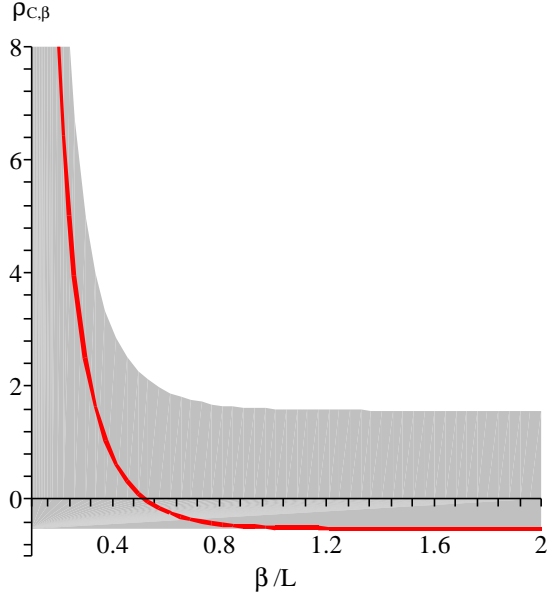}}
\end{center}
\caption{\label{fig:thermal} A graph showing the energy density $\rho_{\Cb,\beta}$
in units of $L^{-2}$ of the
thermal equilibrium state at temperature $\beta^{-1}$ on a cylinder of
circumference $L$. The shaded region indicates the range permitted by
the upper and lower bounds obtained from QEIs, illustrating Eq.~\eqref{eq:thermal_bounds}.}
\end{figure}

Because the energy density is in fact negative, the upper bound was not
particularly enlightening in this example. However the situation is
different for thermal equilibrium states. Let $\omega_{\Cb,\beta}$ be
the thermal equilibrium (KMS) state at inverse temperature $\beta$,
relative to the static time translations. The stress-energy
tensor is again diagonal
\begin{equation}
\langle T_{\Cb\,ab} \rangle_{\omega_{\Cb,\beta}}(x) = \rho_{\Cb,\beta}
\left( \begin{array}{cc}
1 & 0 \\ 0 & 1\\
\end{array}\right),
\end{equation}
where
\begin{equation}
\rho_{\Cb,\beta} = -\frac{\pi}{6L^2} + \frac{\pi}{L^2}\sum_{n=1}^\infty
\csch^2 \frac{n\pi\beta}{L}
\end{equation}
see, e.g., Sec.~4.2 of~\cite{Brl&Dv}. By our general theory, these
states should be constrained by the same lower bound as before, and this
is evidently true, because the series contribution to $\rho_{\Cb,\beta}$
is clearly positive. The upper bound depends on the temperature:
\begin{equation}
\rho_{\Cb,\beta} \le
\frac{\QQ^{\rm weak}_{\Cb}(\fs,\omega_{\Cb,\beta})}{\int_{0}^{\tau_0} g(\tau)^2\,d\tau}
\end{equation}
for any $g\in\tCoinX{(0,\tau_0);\RR}$. In Appendix~\ref{sec:QI_cyl} we obtain
the estimate
\begin{equation}
\QQ^{\rm weak}_{\Cb}(\fs,\omega_{\Cb,\beta}) \le
\frac{\QQ^{\rm weak}_{\Cb}(\fs,\omega_{\Cb})}{1-e^{-2\pi\beta/L}} + 
\frac{\pi e^{\pi\beta/L}}{2L^2\sinh^3\pi\beta/L}\int_{-\infty}^\infty
|g(\tau)|^2\,d\tau 
\label{eq:thermal_estimate_2d_cyl}
\end{equation}
and we may now immediately optimise over $g$ using our result for the
ground state to obtain
\begin{equation}
-\frac{\pi}{6L^2} \le \rho_{\Cb,\beta} \le \frac{\pi}{2L^2(1-e^{-2\pi\beta/L})}
+ \frac{\pi e^{\pi\beta/L}}{2L^2\sinh^3\pi\beta/L}
\label{eq:thermal_bounds}
\end{equation}
As shown in Fig.~\ref{fig:thermal} this is consistent with the known value of
$\rho_{\Cb,\beta}$. 

\subsection{Spatial topology $\RR^{3-j}\times\TT^j$, $j=1,2,3$}

Let us now consider various quotients of four-dimensional Minkowksi
space by subgroups of the group of spatial translations. To begin, 
consider the quotient of four-dimensional Minkowski space, with
inertial coordinates $(t,x,y,z)$ by the spatial
translation subgroup $(t,x,y,z)\mapsto (t,x,y,z+nL_1)$ ($n\in\ZZ$) for some
fixed periodicity length $L_1>0$. We will
denote the resulting spacetime by $\Nb_1$, and consider the ground state
$\omega_{\Nb_1}$, which has a nonzero Casimir
vacuum stress-energy tensor. 
A calculation using the method of images and the Minkowski  
space vacuum two-point function yields the renormalized
vacuum stress-tensor for the massless scalar field in this spacetime, 
(see, e.g.,~\cite{deW&Ha&Is79})
\begin{equation}
\expt{T_{\Nb_1\,ab}}{\omega_{\Nb_1}} = -\frac{\pi^2}{90 L_1^4}
\,{\rm diag}[\, 1,\, -1, \, -1,\, 3 ].
\end{equation}
We will now show that this is consistent with the lower bound arising
from the QEIs. To this end, let $\gamma(\tau)=(\tau,x_0)$ for some
fixed $x_0\in \RR^2\times\TT$. Then the double cone ${\rm
int}\,\left(J^+(\gamma(0))\cap J^-(\gamma(L_1))\right)$ is a c.e.g.h.s.\ of $\Nb_1$
which is causally isometric to a double cone in Minkowski space. Thus
the portion of $\gamma$ parametrised by $(0,L_1)$ meets the hypotheses
of Prop.~\ref{Prop:emaxmassive} and we have
\begin{equation}
\sup_\gamma \langle T_{\Nb_1\,00}\rangle_\omega \ge -\frac{C_4}{L_1^4}
\label{eq:torus_bound}
\end{equation}
for any Hadamard state $\omega$ of the Klein--Gordon field. In particular, the
state $\omega_{\Nb_1}$ obeys this bound, as $\pi^2/90\approx 0.109662<
C_4$. In fact the energy density is about thirty times smaller than
the QEI bound in this case. 

Thus the QEI bounds can be rather weak. But this is necessary, as can be
seen from the next examples, in which the same lower bound must
constrain a more negative energy density. Consider the spacetime
$\Nb_2\cong \RR\times\RR\times \TT^2$, which may be obtained by quotienting $\Nb_1$ by the translation
group $(t,x,y,z)\mapsto (t,x,y+nL_2,z)$ ($n\in\ZZ$) for some nonnegative
$L_2$, which, without loss of generality, we take to be no less than $L_1$. 
Because $L_2\ge L_1$ we may apply Prop.~\ref{Prop:emaxmassive} to a double
cone of the same size as before, so the lower bound is unchanged.
However the stress tensor is now~\cite{deW&Ha&Is79}
\begin{equation}
\expt{T_{\Nb_2\,ab}}{\omega_{\Nb_2}} = -\frac{1}{2\pi^2
L_1^4}\sum_{(m,n)\in \ZZ^2\backslash\{0\}} \frac{1}{(m^2+n^2)^2}\,{\rm diag}[\, 1,\, -1, \, 1,\, 1 ]
\end{equation}
in the special case $L_2=L_1$. The sum can no longer be given in closed
form, but numerically the overall prefactor (equal to the energy
density on the worldline $\tau\mapsto(\tau,x_0)$) is given
in~\cite{deW&Ha&Is79} as $-0.305/L_1^4$. This is still consistent with
Eq.~\eqref{eq:torus_bound}, with energy density now only around ten times
smaller than the bound. 

In exactly the same way we may quotient $\Nb_2$ by the translation
subgroup $(t,x,y,z)\mapsto (t,x+nL_3,y,z)$ ($n\in\ZZ$), thereby forming
$\Nb_3\cong \RR\times \TT^3$. If we again suppose that $L_1\le L_2\le L_3$, then the bound 
Eq.~\eqref{eq:torus_bound} still applies to the ground state on this
spacetime. (Since this spacetime supports normalisable zero modes for
the massless scalar field, one must regard this as a state on the algebra of derivatives of the field,
much as for massless fields in two-dimensions). On the other hand, the
stress-energy tensor in the natural ground state is
\begin{equation}
\expt{T_{\Nb_3\,ab}}{\omega_{\Nb_3}} = -\frac{1}{2\pi^2
L_1^4}\sum_{(l,m,n)\in \ZZ^3\backslash\{0\}} \frac{1}{(l^2+m^2+n^2)^2}
\,{\rm diag}\left[\, 1,\, \frac{1}{3}, \, \frac{1}{3},\, \frac{1}{3} \right]
\end{equation}
in the special case $L_1=L_2=L_3$. The energy density along
$\tau\mapsto(\tau,x_0)$ in this case is
numerically computed to be $0.838/L_1^4$, which is again consistent with
the QEI constraint Eq.~\eqref{eq:torus_bound}, which is now weaker by a factor of less than $4$. 

Let us note that the massless QEI bound also provides a lower bound on
the ground state energy densities of {\em massive} scalar fields in
these spacetimes. Consistency here is seen from the fact that the mass
diminishes the magnitude of the energy density~\cite{Tan&His95} (note
the misprints in~\cite{Tan&His95} noted in~\cite{Lan05} which do not,
however, affect the final result).

\subsection{Misner Universe}

Our third example concerns (the globally hyperbolic portion of) the Misner
universe $\Ub$; namely, the quotient of $(0,\infty)\times\RR^3$ with metric
\begin{equation}
ds^2 = dt^2 - t^2 \left(dx^1\right)^2 -\left(dx^2\right)^2-\left(dx^3\right)^2\,,
\label{eq:metric_misner}
\end{equation}
by the translation group $(t, x^1, x^2, x^3)\mapsto (t, x^1+na, x^2,
x^3)$ ($n\in\ZZ$) for some constant $a>0$. That is, the $x^1$
coordinate has been compactified onto a circle. We restrict to $t>0$
to avoid the closed null geodesics which would appear at $t=0$ and the
closed timelike curves appearing for $t<0$. Under the coordinate
transformation
\begin{equation}
y^0 = t \cosh(x^1), \qquad y^1 = t \sinh(x^1), \qquad y^2=x^2, \qquad y^3=x^3,
\end{equation}
we may, equivalently, regard Misner space as the wedge
$y^0>|y^1|$ of Minkowski spacetime with the points $(y^0,y^1,y^2,y^3)$
and $(y^0\cosh(na)+y^1\sinh(na), y^1\cosh(na)+y^0\sinh(na),y^2,y^3)$
identified for each $n\in\ZZ$.


\begin{figure}
\begin{center}
\includegraphics{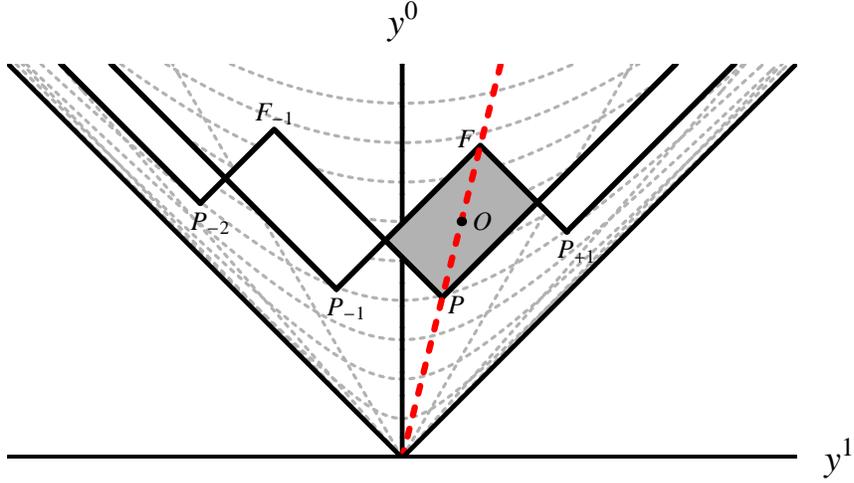} 
\end{center}
\caption{\label{fig:misner} The covering space of the Misner universe
is the wedge $y^0 > |y^1|$ of Minkowski spacetime, shown here
in cross section in the $y^0$-$y^1$ coordinates.  Points in the covering
space are identified, as described in the text, along the background
hyperbol\ae.  Also shown is a stationary geodesic in the Misner universe,
which in the covering space is a constant velocity observer (dashed line).  
For such an observer, the largest causal diamond for a given center point
that is isomorphic to a subset of Minkowski space is shown in gray.  
Identified images of this diamond are also shown in white.}
\end{figure}


Define a curve $\gamma(\tau)=(\tau,\xb_o)$ in the original coordinates,
for some constant $x_0\in \TT\times\RR^2$. This is a timelike
geodesic, with velocity $u^a=(1,\boldsymbol{0})$. In the Minkowski
space cover, this worldline is given by
\begin{equation}
\gamma(\tau) = \left(\begin{array}{c}\tau \cosh (x_o^1)\\[3pt]
\tau \sinh (x_o^1)\\[3pt]
x_o^2\\[3pt] x_o^3 \end{array}\right)
\qquad\mbox{ with }\qquad
u^a(\tau) = \left(\begin{array}{c} \cosh (x_o^1)\\[3pt] \sinh (x_o^1)\\[3pt]
0 \\[3pt] 0  \end{array}\right),
\end{equation}
which is a constant velocity geodesic, as shown by the bold dashed line in 
Figure~\ref{fig:misner}. Let us consider a portion of this curve,
running between $P=\gamma(\tau_P)$ and $F=\gamma(\tau_F)$, which is
such that ${\rm int}(J^+(P)\cap J^-(F))$ is a c.e.g.h.s.\ of Misner
space isometric to a double cone in Minkowski space. Assuming that
this region is maximal, it must be that the geodesic joining the
$n=+1$ image $P_{+1}$ of $P$ to $F$ is null. Setting
$\tau_{O}=(\tau_F+\tau_P)/2$ and $L=(\tau_F-\tau_P)/2$, this yields
the condition
\begin{equation}
\left[\left(\tau_O+\frac{L}{2}\right)\cosh(x_o^1) -
\left(\tau_O-\frac{L}{2}\right)\cosh(x_o^1+a)\right]^2-
\left[\left(\tau_O+\frac{L}{2}\right)\sinh(x_o^1) -
\left(\tau_O-\frac{L}{2}\right)\sinh(x_o^1+a)\right]^2 =0,
\end{equation} 
which entails
\begin{equation}
L = 2 \tau_O \tanh\left(\frac{a}{2}\right).
\end{equation} 
This is the largest double cone of this type, centered on $O$, in
which an observer cannot detect the compactified nature of the $x^1$-direction. 
We may therefore apply Prop.~\ref{Prop:emaxmassive} to the portion
$\gamma_{PF}$ of $\gamma$ lying between $P$ and $F$. This gives
\begin{equation}
\sup_{\gamma_{PF}} \langle T_{\Ub\,ab} u^a u^b\rangle_\omega  \ge -\frac{C_4}{L^4} 
= -\frac{C_4}{(2\tau_O\tanh(a/2))^4}
\end{equation}
for any Hadamard state $\omega$ on Misner space.
In particular, an energy density $\rho(\tau)=\langle
T_{\Ub\,ab}u^au^b\rangle_\omega(\gamma(\tau))$ of the form
$\rho(\tau)=-C/\tau^4$, for which $\sup_\gamma \rho=-C/\tau_F^4$, 
would be subject to the constraint
\begin{equation}
C \le \frac{C_4}{16}\left(2+\coth(a/2)\right)^4. \label{eq:Misner_first_go}
\end{equation}
By adapting the eigenvalue method, we may obtain a better bound. Let us suppose
that $\rho$ obeys
\begin{equation}
\rho(\tau)\le \frac{K}{16\pi^2\tau^4}
\end{equation}
on $I=(\tau_O-L/2,\tau_0+L/2)$. Then by exactly the same arguments as
in Sec.~\ref{sect:GA_inertial}, we may deduce
\begin{equation}
K \geq -\inf_g \frac{\int |g''(\tau)|^2 d\tau}
{\int  \tau^{-4} |g(\tau)|^2 d\tau}.
\label{Eq:misnerQI}
\end{equation}
where the $\tau^{-4}$ in the denominator comes from the form of
$\rho$, and the infimum is taken over all $g\in\CoinX{I}$. The
denominator can be reinterpreted as the norm of $g$ in
$L^2(I,\tau^{-4}d\tau)$. Integrating by parts twice, we may rewrite
the numerator as $-\ip{g}{Lg}$ where the inner product is that of
$L^2(I,\tau^{-4}d\tau)$ and $L$ is defined on $\CoinX{I}$ by
\begin{equation}
(Lg)(\tau) = \tau^4g^{\prime\prime\prime\prime}(\tau)
\end{equation}
and is symmetric, i.e., $\ip{h}{Lg}= \ip{Lh}{g}$ for all
$g,h\in\CoinX{I}$. The minimisation problem is then solved by finding
the lowest spectral point of the Friedrichs extension of $L$. It may
be shown that the Friedrichs extension again amounts to the imposition
of Dirichlet boundary conditions $g=g'=0$ on $\partial
I$ \footnote{The Friedrichs extension has a domain contained in the
closure of $\CoinX{I}$ in the norm $\|g\|_{+1}^2 =
\ip{g}{g}+\ip{g}{Lg}$, which is equivalent to the norm of the Sobolev
space $W_0^2(I)$ since $\tau^{-4}$ is bounded and bounded away from
zero on $I$. Accordingly, the closure of $\CoinX{I}$ is precisely
$W_0^2(I)$ and the desired domain lies in this Sobolev space, all
elements of which obey $g=g'=0$ on $\partial I$.}, 
and the
problem now reduces to the study of the ODE
\begin{equation}
g''''(\tau)-\frac{\lambda}{\tau^4}g(\tau)=0.
\end{equation}
Again we wish to determine the minimum eigenvalue $\lambda$ for eigensolutions
that satisfy the boundary conditions. The substitution
$g(\tau)=h(\frac{1}{2}\ln(\tau))$ converts the equation to a constant
coefficient linear equation, and one may determine the general
solution (e.g., using {\em Mathematica}) as
\begin{equation}
h(l)=c_1e^{3l}\cos\left(l\sqrt{4\sqrt{\lambda+1}-5}\right)
+c_2 e^{3l}\sin\left(l\sqrt{4\sqrt{\lambda+1}-5}\right)
+c_3 e^{\left(3+\sqrt{4\sqrt{\lambda+1}+5}\right)l}
+c_4 e^{\left(3-\sqrt{4\sqrt{\lambda+1}+5}\right)l},
\end{equation}
where $( c_1, c_2, c_3, c_4 )$ are constants. Imposing three of the boundary 
conditions fixes three of the constants in terms of the fourth, which serves
as an overall magnitude for the test function.  The fourth boundary condition
can then be used to determine the eigenvalues.  A somewhat involved
calculation leads to the transcendental equation to determine
$\lambda$ implicitly in terms of $a$:
\begin{equation}
\frac{\sqrt{16\lambda-9}}{5} =\frac{\sin\left(\frac{a}{2}\sqrt{4
\sqrt{\lambda+1}-5}\right)\sinh\left(\frac{a}{2}\sqrt{4
\sqrt{\lambda+1}+5}\right)}{\cos\left(\frac{a}{2}\sqrt{4
\sqrt{\lambda+1}-5}\right)\cosh\left(\frac{a}{2}\sqrt{4
\sqrt{\lambda+1}+5}\right)-1}.
\end{equation}
We denote $\lambda$, so determined, as $\lambda(a)$; this constrains
our original value $K$ by
\begin{equation}
K\geq -\lambda(a).
\end{equation}

\begin{figure}
\begin{center}
\rotatebox{0}{\scalebox{.75}{
\includegraphics{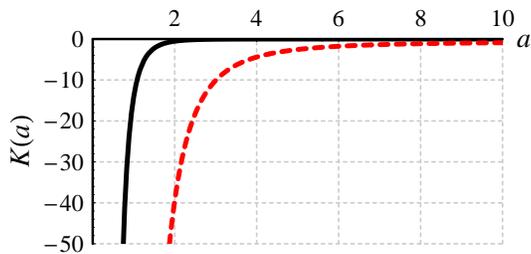} 
}}
\end{center}
\caption{\label{fig:misner_bnd} The numerical factor $K(a)$ for the
vacuum stress tensor in the Misner universe (solid line) plotted for
a range of the closure scale $a$.  Also shown is the lower bounds
from $-\lambda(a)$ (dashed line).   
The lower bound from Eq.~\eqref{eq:Misner_first_go} is so weak 
(it asymptotically approaches $\sim -2500$ from below for large $a$)
that it is not included to preserve detail in the figure above.
}
\end{figure}

Our interest in energy densities proportional to $\tau^{-4}$ stems
from the state constructed by Hiscock and Konkowski
\cite{Hi&K82}. This quasifree state, which we denote $\omega_\Ub$, is obtained by applying the
method of images to the Minkowski space two-point function in the
wedge $y^0>|y^1|$, and then carrying it back to the original Misner
coordinates to find the renormalized vacuum expectation value of the 
stress-tensor. Hiscock and Konkowski considered the conformally coupled 
scalar field, but their calculations can be easily reproduced in the
minimally-coupled case, to yield
\begin{equation}
\expt{T_{\Ub\,ab}}{\omega_\Ub}(t) = \frac{K( a)}{16\pi^2\,t^4}\,
\mbox{diag}\left[ 1, 3t^2, -1, -1\right],
\end{equation}
where
\begin{equation}
K(a)=-\sum_{n=1}^\infty \csch^4\left(\frac{na}{2}\right)
\end{equation}
is a negative constant depending on the $x^1$-period $a$
\footnote{For a scalar field with arbitrary curvature coupling constant
$\varepsilon$, replaced the numerical coefficient $K(a)$ in the stress-tensor
with $K_\varepsilon(a)= -\sum_{n=1}^\infty \left[ 
\csch^4(na/2) +4\varepsilon\, 
\csch^2(na/2)\right]$.}. Both the coefficient $K(a)$ and the numerical 
evaluation of the lower bound $-\lambda(a)$ are plotted in 
Fig.~\ref{fig:misner_bnd}.  It is obvious that $K(a)$, and thus the 
energy density obey the QEI constraint for all values of $a$. The bound
Eq.~\eqref{eq:Misner_first_go} is still weaker. 


\subsection{Rindler spacetime}

The Rindler spacetime $\Rb$ is the ``right wedge'' of Minkowski space,
i.e., the region
$\left\{ (t,x,y,z)\in\RR^4 : \mbox{ s.t. } x >|t|\right\}$ in inertial
coordinates $(t,x,y,z)$. We may also make the
coordinate transformation
\begin{equation}
\begin{array}{ccl}
t = \xi \sinh(\eta), &\qquad\qquad& y = y,\\[4pt]
x = \xi \cosh(\eta), &\qquad\qquad& z = z,
\end{array}
\end{equation}
to obtain the metric in the form
\begin{equation}
ds^2=\xi^2 d{\eta}^2-d{\xi}^2-d{y}^2-d{z}^2,
\end{equation}
with coordinate ranges $\eta,y,z\in\RR$, $\xi \in (0,\infty)$.  
Lines of constant $\xi$, when mapped
into Minkowski space, are worldlines for observers undergoing constant proper
acceleration $\alpha=\xi^{-1}$.  Rindler spacetime is static with respect to $\eta$
(corresponding to Lorentz invariance in the $xt$ plane) and is
invariant under Euclidean transformations of the
$yz$ plane. 

Clearly any line of constant $\xi$ meets the conditions of
Prop.~\ref{prop:emax_4d_accel}
and we may immediately read off that any static Hadamard state $\omega$
on $\Rb$ must obey
\begin{equation}
\expt{T_{\Rb\,ab}u^a u^b}{\omega}(\eta,\xi,y,z) \ge -\frac{11}{480\pi^2\xi^4}
\end{equation}
where $u^a$ is the unit vector parallel to $\partial/\partial\eta$. In
particular, this provides a constraint on the energy density $\rho_\Rb =
\langle T_{\Rb\,ab}u^au^b\rangle_{\omega_{\Rb}}$ in the ground state
$\omega_\Rb$ (which is Hadamard). This may also be computed exactly: 
it was first computed for the conformally coupled scalar field
by Candelas and Deutsch~\cite{C&De77} and one can easily generalize their results to the
minimally coupled scalar field to obtain~\footnote{To generalize the
result to a scalar field
with {\em arbitrary} curvature coupling
constant $\zeta$, replace the $11$ in the numerator with
$(11-60\zeta)$.}
\begin{equation}
\rho_\Rb =  -\frac{11}{480\pi^2\xi_o^4},
\end{equation} 
which is exactly the lower bound given 
above. Thus, remarkably, the Rindler ground state saturates the QEI
constraints, which were obtained using local covariance and the
Minkowski vacuum, and nowhere involved $\omega_\Rb$.


\begin{figure}
\begin{center}
\includegraphics{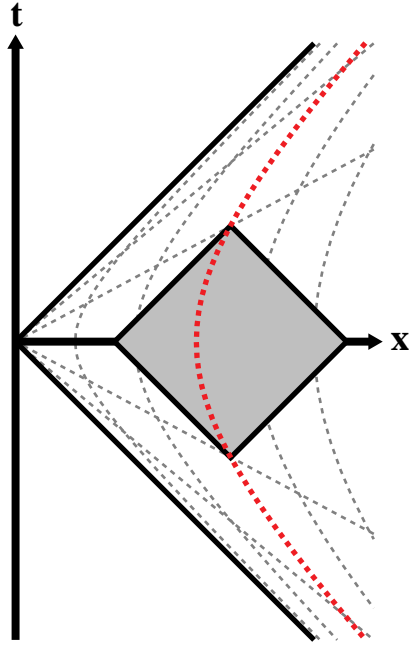} 
\end{center}
\caption{\label{fig:rindler} Diagram showing Rindler spacetime
(with the two perpendicular space dimensions suppressed) embedded
into Minkowski
spacetime.  The dashed hyperbolic line, the worldline of a constantly
accelerating observer, is the image of a constant $\xi$ observer's
worldline in Rindler coordinates.  The grey diamond is a causal region
that can be isometrically identified between the two ``different''
coordinate systems.}
\end{figure}

Let us also examine how an upper bound might be obtained. Let
$\gamma(\tau)=(\tau/\xi_o,\xi_o,0,0)$ in $(\eta,\xi,y,z)$ coordinates
and set $\us=\dot{\gamma}$ as usual. We consider sampling along
$\gamma$, with sampling tensors of form
\begin{equation}
\fs(\ts) = \int t_{ab}|_{\gamma(\tau)} u^au^b g(\tau)^2\,d\tau
\end{equation}
for $g\in\tCoinX{\RR;\RR}$. Since the energy density is constant along
$\gamma$, the upper bound of 
Cor.~\ref{Cor:diffMink} gives
\begin{equation}
\rho_\Rb\int g(\tau)^2\,d\tau = \langle \Ts_\Rb(\fs)\rangle_{\omega_\Rb}
\le \QQ^{\rm weak}_\Rb(\fs,\omega_\Rb)\,.
\end{equation}
The right-hand side can be read off from the
difference QEI derived by Pfenning \cite{Pfen02} for
the electromagnetic field, because the corresponding bound for the scalar
field is exactly half of the electromagnetic expression 
\footnote{Note that
the weight function in~\cite{Pfen02} was parametrised in terms of
$\eta$, rather than proper time $\tau$: our $g(\tau)$ is related to
the $f(\eta)$ of~\cite{Pfen02} by
$g(\tau)=\sqrt{f(\tau/\xi_o)/\xi_o}$.}: 
\begin{eqnarray}
\QQ^{\rm weak}_\Rb(\fs,\omega_\Rb)
&=& \frac{1}{16\pi^3}\int_0^\infty  \left|
\widehat{g}(u) \right|^2 \left(u^4+2\xi_o^{-2}u^2\right)du\nonumber\\
&=& \frac{1}{16 \pi^2} \int_{-\infty}^\infty
\left( \left| g''(\tau) \right|^2 + 2\xi_o^{-2} \left| g'(\tau)
\right|^2\right) d\tau\,.
\end{eqnarray}
Next consider scaling the test function, replacing $g$ by
$g_\alpha(\tau)=\alpha^{-1/2}g(\tau/\alpha)$. 
We find, considering the scaling behavior of the above expression,
\begin{equation}
\rho_\Rb \le \frac{1}{16\pi^2}\frac{\int\left(
|g''(\tau)|^2 + 2\alpha^2\xi_o^{-2} |g'(\tau)|^2\right)d\tau}{\alpha^4\int g(\tau)^2 d\tau}
\end{equation}
for which the right hand side vanishes in the limit of
$\alpha\rightarrow\infty$. Thus we find consistency with the known fact
that the expectation value of the 
Rindler ground state is bounded above by zero, i.e. $\rho_\Rb\le 0$.

\section{Summary}

In this paper we have initiated the study of interrelations between
quantum energy inequalities and local covariance. We have formulated
definitions of locally covariant QEIs, and shown that existing QEIs
obey them, modulo small additional restrictions
(Sec.~\ref{sec:QEIs_and_lc}). The main thrust of our work has been
directed at providing {\em a priori} constraints on renormalised energy
densities in locally Minkowskian regions, accomplished in
Sec.~\ref{sec:general_apps}. The simple geometric nature of
these bounds makes them easy to apply in practice, and a number of
future applications are envisaged. In particular, we will discuss
applications to the Casimir effect in a companion paper~\cite{cas_b}; at
the theoretical level, it is possible to place the present discussion in the
categorical language of~\cite{BrFrVe03}, and this will be done elsewhere.
Equally important are the specific calculations reported in
Sec.~\ref{sec:specific_apps}. Here we saw that, in some situations, the
QEI bounds give best-possible constraints on the energy density, and
that typical ground state energy densities are not over-estimated 
by the QEI bound by more than a factor of about $30$ at worst (in the
examples so far studied). 

Finally, although we confined our attention largely to locally
Minkowskian spacetimes in Secs.~\ref{sec:general_apps}
and~\ref{sec:specific_apps}, we emphasise again that other interesting
cases may be studied using our general formalism, as, for example, in
the work of Marecki~\cite{Marecki05} on spacetimes with locally Schwarzschild
subregions. 


\appendix

\section{The locally covariant quantum field theory of a scalar field}
\label{appdx:Locally_Covar_QFT}

In this appendix we describe the construction of the quantised
Klein--Gordon field within the algebraic approach to quantum field
theory, and explain the construction of pulled back states used in  
Sec.~\ref{sect:localcovariance}.

The free scalar field of mass $m\ge 0$ may be quantised on any globally
hyperbolic spacetime $\Mb$ in the sense that one may construct a 
complex unital $*$-algebra $\Af(\Mb)$ whose elements may be interpreted
as  `polynomials in
smeared fields'. A typical element of the algebra is a complex linear
combination of the identity $\II$ and a finite number of terms each of
which is a finite product of a number of objects $\Phi_\Mb(f)$ where $f$ is a
test function (i.e., smooth and compactly supported) on $M$. The algebra
also satisfies a number of relations:
\begin{enumerate}
\item $\Phi_\Mb(\lambda f+\mu g) = \lambda\Phi_\Mb(f)+\mu\Phi_\Mb(g)$
\item $\Phi_\Mb(f)^*=\Phi_\Mb(\overline{f})$
\item $\Phi_\Mb((\Dal+m^2)f) =0$ 
\item $[\Phi_\Mb(f),\Phi_\Mb(g)]= iE_{\Mb}(f,g)\II$
\end{enumerate}
for all test functions $f,g$ on $M$ and complex scalars $\lambda,\mu$, 
where $E_{\Mb}$ is the advanced-minus-retarded fundamental solution to
$\Dal+m^2$ on $\Mb$. The first two axioms are necessary for compatibility
with the idea of $\Phi_\Mb(f)$ as a smeared hermitian field; the third
expresses the field equation in `weak' form; the fourth expresses the
commutation relations. 

Now let $\psi$ be a causal isometric embedding of $\Mb_1$ into
$\Mb_2$. Any test function $f$ on $M_1$ now corresponds to a test
function $\psi_*f$ on $M_2$, defined by $(\psi_* f)(x)=f(\psi^{-1}(x))$ for
$x\in\psi(M_1)$ and $(\psi_*f)(x)=0$ otherwise. We may use this to
define a map $\alpha_\psi$ between $\Af(\Mb_1)$
and $\Af(\Mb_2)$ such that
\begin{enumerate}
\item $\alpha_\psi\II_{\Af(\Mb_1)}=\II_{\Af(\Mb_2)}$
\item $\alpha_\psi(\Phi_{\Mb_1}(f)) = \Phi_{\Mb_2}(\psi_* f)$
for all test functions $f$ on $M_1$
\item $\alpha_\psi$ extends to general elements of $\Af(\Mb_1)$ as a
$*$-homomorphism, i.e., $\alpha_\psi$ is linear and obeys
$\alpha_\psi(AB)=\alpha_\psi(A)\alpha_\psi(B)$ and
$\alpha_\psi(A^*)=\alpha_\psi(A)^*$ for all $A,B\in \Af(\Mb_1)$. 
\end{enumerate}
In the body of the text we have used the notation $\psi_*$ for
$\alpha_\psi$, relying on the context for the appropriate meaning; 
here, it is convenient to distinguish the two maps.
One must check that the last statement is compatible with the axioms
stated above--the only nontrivial one is the commutation relation, where
the causal nature of $\psi$ plays a key role and guarantees that
$\alpha_\psi$ is well-defined. What needs to be proved boils down to
checking that 
\begin{equation}
E_{\Mb_1}(f,g)=E_{\Mb_2}(\psi_*f,\psi_*g)
\end{equation}
for all test functions $f,g$ on $M_1$. This equivalence is proved as
follows. Writing $E^\pm_{\Mb}$ for the advanced ($-$) and retarded $(+)$
Green functions on $\Mb$, $E^\pm_{\Mb_2}\psi_*f$ solves the inhomogeneous
Klein--Gordon equation on $\Mb_2$ with source $\psi_*f$ and support in
$J^\pm_{\Mb_2}(\supp f)$. Because $\psi$ is a causal isometry, the
pull-back $\psi^*E^\pm_{\Mb_2}\psi_*f$ solves the inhomogeneous
Klein--Gordon equation on $\Mb_1$ with source $f$ and support in
$J^\pm_{\Mb_1}(\supp f)$; by uniqueness of solution, we have 
$E_{\Mb_1}^\pm f =\psi^* E_{\Mb_2}^\pm\psi_*f$. 
Accordingly $\psi_* E_{\Mb_1}=E_{\Mb_2}\psi_*$ and the required result follows. 

In the algebraic approach we have been pursuing, a state of the quantum
field on $\Mb$ is a linear map $\omega$ from $\Af(\Mb)$ to the complex
numbers, obeying $\omega(\II)=1$ and $\omega(A^*A)\ge 0$ for any $A\in
\Af(\Mb)$. One interprets $\omega(A)$ as the expectation value of
observable $A$ in
state $\omega$. In particular, each state yields a hierarchy of $n$-point functions,
i.e., maps of the form
\begin{equation}
(f_1,\ldots,f_n) \mapsto \omega(\Phi_\Mb(f_1)\cdots\Phi_\Mb(f_n))\,;
\end{equation}
we will restrict attention to those states whose corresponding $n$-point
functions are distributions. A state $\omega$ is Hadamard if its
two-point function has a particular singular structure which is
determined by the local metric and causal properties of the spacetime.
Note that none of the structure introduced so far invokes any particular Hilbert space
representation of the theory. 

Now suppose again that $\psi$ is a causal isometric embedding of $\Mb_1$ into
$\Mb_2$ and let $\omega_2$ be a state on $\Af(\Mb_2)$. We obtain
a state $\omega_1$ on $\Af(\Mb_2)$ by
\begin{equation}
\omega_1(A)=\omega_2(\alpha_\psi(A))
\end{equation}
for any $A\in \Af(\Mb_1)$; that is,
$\omega_1=\alpha_\psi^*\omega_2$, where $\alpha_\psi^*$ is the dual map
to $\alpha_\psi$ (in the body of the text, we have written $\psi^*$
for $\alpha_\psi^*$). The $n$-point functions are therefore related by
\begin{equation}
\omega_1(\Phi_{\Mb_1}(f_1)\cdots\Phi_{\Mb_1}(f_n))=
\omega_2(\alpha_\psi(\Phi_{\Mb_1}(f_1)\cdots\Phi_{\Mb_1}(f_n)))=
\omega_2(\Phi_{\Mb_2}(\psi_* f_1)\cdots\Phi_{\Mb_2}(\psi_* f_n))\,.
\end{equation}
It is useful to write this in `unsmeared' notation. Let $w_j^{(n)}$ be
the $n$-point functions of $\omega_j$. Then the last equation becomes
\begin{eqnarray}
\lefteqn{\int_{M_1} d{\rm vol}_{\gb_1}(x_1)\cdots\int_{M_1}d{\rm vol}_{\gb_1}(x_n)\,
w_1^{(n)}(x_1,\ldots,x_n)f_1(x_1)\cdots f_n(x_n)} &&\nonumber\\
&=&\int_{M_2} d{\rm vol}_{\gb_2}(y_1)\cdots\int_{M_2}d{\rm vol}_{\gb_2}(y_n)\,
w_2^{(n)}(y_1,\ldots,y_n)\psi_*f_1(y_1)\cdots \psi_*f_n(y_n)\nonumber\\
&=&\int_{M_1^{\times n}} d{\rm vol}_{\gb_1}(x_1)\cdots\,d{\rm vol}_{\gb_1}(x_n)
w_1^{(n)}(\psi(x_1),\ldots,\psi(x_n))f_1(x_1)\cdots f_n(x_n)\,,
\end{eqnarray}
where the change of variables employed in the last step is justified by
the fact that $\psi$ is an isometry. 
As this holds for all choices of $f_k$, we may deduce that
\begin{equation}
w_1^{(n)}(x_1,\ldots,x_n) = w_2^{(n)}(\psi(x_1),\ldots,\psi(x_n))\,;
\end{equation}
that is, the $n$-point functions of $\omega_1$ are the pull-backs by
$\psi$ of those of $\omega_2$. It follows that if $\omega_2$ is Hadamard
then so too is $\omega_1$, because the two-point function is simply
pulled back under $\psi$ and the Hadamard series is constructed from the
local causal and metric structure which is preserved under $\psi$. Since
the stress-energy tensor is renormalised by subtracting the first few
terms of the Hadamard series from the two point function, and then
taking suitable derivatives before taking the coincidence limit (making
a further locally constructed correction to ensure conservation of the
stress tensor), we have the following important consequence, which we
isolate as a theorem.

\begin{Thm} Suppose $\psi$ is a causal isometric embedding of
globally hyperbolic spacetime $\Mb_1$ 
in a globally hyperbolic spacetime $\Mb_2$. Any Hadamard state of
the massive Klein--Gordon quantum field on $\Mb_2$ induces a
Hadamard state of the same theory on $\Mb_1$, whose $n$-point
functions and renormalised expected stress-energy tensor are the pull-backs by
$\psi$ of the corresponding quantities on $\Mb_2$. 
\end{Thm}

This fits in with the principle that one should not be able to tell,
by local experiments, whether one is in $\Mb_1$ or its image within 
the larger spacetime $\Mb_2$. It also justifies us in the abuse of
notation perpetrated in Sec.~\ref{sect:localcovariance}, where we wrote $\psi^*$ in place of
$\alpha_\psi^*$, and (dually) $\psi_*$ in place of $\alpha_\psi$. 

Let us conclude by briefly describing more of the structure set out by~\cite{BrFrVe03}.
The key is the observation that the 
globally hyperbolic spacetimes of given dimension form the objects of a
category in which the morphisms are causal isometric embeddings. One may
also consider a category of unital $*$-algebras with injective unit-preserving
$*$-homomorphisms as morphisms. The association of a globally hyperbolic
spacetime $\Mb$ with the corresponding algebra $\Af(\Mb)$ is then
shown to be a covariant functor between these categories and gives a
precise meaning to the notion of `the same field theory on different
spacetimes' (and the same would be true even for theories not
necessarily described in terms of a Lagrangian). A similar functorial
description may be given to the association of the state space of the
theory, and quantum fields are reinterpreted as natural transformations
between functors. We refer the reader to~\cite{BrFrVe03} for full details.

\section{Derivation of Scalar Field Quantum Weak Energy Inequality
in the Cylinder Spacetime}\label{sec:QI_cyl}

In this appendix, we calculate quantum weak energy inequalities for the
massless, minimally--coupled real scalar field in the two dimensional
cylinder spacetime relative to the ground and thermal equilibrium states.
We use the notation of Sec.~\ref{sec:2D_cyl}.

The KMS state $\omega_{\Cb,\beta}$ at inverse temperature $\beta$ has
two-point function (see, e.g., Eq.~(2.43) of~\cite{Fu&Ru87})
\begin{equation}
w_{2,\beta}(x ,x') =
\frac{1}{2L} \sum_{n\in\ZZ\backslash\{0\}} \frac{e^{ik_n(z-z')}}{\omega_n(1-e^{-\beta\omega_n})}\,
\left(e^{-i\omega(t-t')}+ e^{-\beta\omega_n}e^{i\omega(t-t')}\right),
\end{equation}
where $k_n= 2\pi n/L$ and $\omega_n=|k_n|$, and the sum converges in the
distributional sense (i.e., after smearing each term with test
functions, the resulting series converges and its sum depends
continuously on the test functions). We exclude the zero mode $n=0$ as
usual, regarding $\omega_{\Cb,\beta}$ as a state on the derivative fields. 
The two-point function of the ground state $\omega_\Cb$ is
obtained as the zero temperature ($\beta\to\infty$) limit of this
expression. We will be interested in the static curve
$\gamma(\tau)=(\tau,0)$, and employ the tetrad $e_0=\partial/\partial
t$, $e_1=\partial/\partial z$, which is invariant under Fermi--Walker
transport along $\gamma$. 

Following the procedure of Sec.~\ref{sect:lcdQEIs_examples}, we
find
\begin{equation}
Q_{\gamma,\omega_{\Cb,\beta}}(u) = \frac{1}{2\pi^2} \int_{(-\infty,u)} dv\,
\widehat{T}_{\gamma,\omega_{\Cb,\beta}}(v) 
\end{equation}
and
\begin{equation}
T_{\gamma,\omega_{\Cb,\beta}}(\sigma) 
=\frac{1}{L}\sum_{n=1}^\infty
\frac{\omega_{n}}{1-e^{-\beta\omega_n}}\left(
e^{-i\omega_n\sigma} + e^{-\beta\omega_n}e^{i\omega_n\sigma}\right)\,.
\end{equation}
Taking the Fourier transform, we have 
\begin{equation}
\widehat{T}_{\gamma,\omega_{\Cb,\beta}}(v) 
=\frac{2\pi}{L}\sum_{n=1}^\infty
\frac{\omega_{n}}{1-e^{-\beta\omega_n}}\left(
\delta(v-\omega_n) + e^{-\beta\omega_n}\delta(v+\omega_n)\right)
\end{equation}
and therefore obtain
\begin{equation}
Q_{\gamma,\omega_{\Cb,\beta}}(u) = 
\frac{1}{\pi L} \left\{\sum_{\stack{n\in\NN}{{\rm s.t.}\omega_n<u}}
\frac{\omega_n}{1-e^{-\beta\omega_n}}
+ \sum_{\stack{n\in\NN}{{\rm s.t.}-\omega_n<u}}
\frac{\omega_n e^{-\beta\omega_n}}{1-e^{-\beta\omega_n}} \right\}\,.
\end{equation}
Note that $Q_{\gamma,\omega_{\Cb,\beta}}$ is supported on $\RR^+$ in the $\beta=\infty$ case, but otherwise
on the whole of $\RR$, albeit exponentially suppressed on the negative
half line. 

To arrive at convenient QEI bounds we estimate
$Q_{\gamma,\omega_{\Cb,\beta}}$ by
\begin{equation}
Q_{\gamma,\omega_{\Cb,\beta}}(u)\le Q_{\gamma,\omega_{\Cb,\beta}}(0) + 
\frac{\vartheta(u)}{\pi L(1-e^{-2\pi\beta/L})}\sum_{\stack{n\in\NN}{{\rm
s.t.}\omega_n<u}} \omega_n
\end{equation}
where $\vartheta$ is the Heaviside step function. To see that this
estimate is valid, we note that $Q_{\gamma,\omega_{\Cb,\beta}}(u)$ is
clearly increasing on the negative half-line, so it is valid to bound it
by $Q_{\gamma,\omega_{\Cb,\beta}}(0)$ on $u\le 0$; the second term in the
estimate arises by noting that
$(1-e^{-\beta\omega_n})^{-1}\le(1-e^{-2\pi\beta/L})^{-1}$ for all $n$.
Using this estimate again, we also have
\begin{equation}
Q_{\gamma,\omega_{\Cb,\beta}}(0)= 
\sum_{n\in\NN}
\frac{\omega_n e^{-\beta\omega_n}}{1-e^{-\beta\omega_n}}
\le \frac{1}{\pi L(1-e^{-2\pi\beta/L})}\sum_{n=1}^\infty \frac{2\pi
n}{L} e^{-2\pi\beta n/L} = \frac{e^{\pi\beta/L}}{4L^2\sinh^3\pi\beta/L}\,,
\end{equation}
while, for $u> 0$
\begin{equation}
\frac{1}{\pi L}\sum_{\stack{n\in\NN}{{\rm
s.t.}\omega_n<u}} \omega_n  = \frac{1}{L^2} \tilde{n}(\tilde{n}+1) \le \frac{2}{L^2} \tilde{n}^2
\le \frac{u^2}{2\pi^2}\,,
\end{equation}
where $\tilde{n}$ is the greatest integer {\em strictly} less than
$uL/(2\pi)$. Thus we have the estimate
\begin{equation}
Q_{\gamma,\omega_{\Cb,\beta}}(u)\le
\frac{e^{\pi\beta/L}}{4L^2\sinh^3\pi\beta/L}
+ \frac{\vartheta(u)u^2}{2\pi^2(1-e^{-2\pi\beta/L})}\,.
\end{equation}
It then follows that, in the notation of Sec.~\ref{sec:2D_cyl},
\begin{eqnarray}
\QQ^{\rm weak}_\Cb(\fs,\omega_{\Cb,\beta}) &=& \int_{-\infty}^\infty du \,|\widehat{g}(u)|^2
Q_{\gamma,\omega_{\Cb,\beta}}(u) \nonumber\\
&\le&
\frac{e^{\pi\beta/L}}{4L^2\sinh^3\pi\beta/L}\int_{-\infty}^0 |\widehat{g}(u)|^2\,du +
\frac{1}{2\pi^2(1-e^{-2\pi\beta/L})}\int_0^\infty  u^2 |\widehat{g}(u)|^2
\,du \\
&\le& \frac{\pi e^{\pi\beta/L}}{2L^2\sinh^3\pi\beta/L}\int_{-\infty}^\infty
|g(\tau)|^2\,d\tau +
\frac{1}{2\pi(1-e^{-2\pi\beta/L})}\int_{-\infty}^\infty
|g'(\tau)|^2\,d\tau 
\end{eqnarray}
where we have used Parseval's theorem and the fact that
$|\widehat{g}(u)|$ is even for real-valued $g$ to convert the integral over
$\RR^+$ into one over $\RR$. For the ground state, of course, this yields
\begin{equation}
\QQ^{\rm weak}_\Cb(\fs,\omega_{\Cb}) \le 
\frac{1}{2\pi}\int_{-\infty}^\infty
|g'(\tau)|^2\,d\tau 
\end{equation}
and Eq.~\eqref{eq:thermal_estimate_2d_cyl} follows immediately. Although our estimates are not very sharp,
they have led to a very simple quantum inequality.
In fact, for the ground state, this inequality is only $3$ times less restrictive than the optimal
quantum inequality bound found in two dimensional Minkowski spacetime.

\section{A lemma concerning smooth functions} \label{appx:smooth}

Recall from Sec.~\ref{sect:lcdQEIs_examples} that we define 
$\tCoinX{I;\RR}$ to be the set of smooth compactly supported
real-valued functions $g$ on $I$ whose support is connected and which have
no zeros of infinite order in the interior of that support (equivalently, $g$ has no zeros
in $(\inf\supp g,\sup\supp g)$ of infinite order). Our aim is to prove
the following result. 
\begin{Lem} Let $g\in\CoinX{I;\RR}$ and choose any
$\chi\in\tCoinX{I;\RR}$ with $\chi=1$ on $\supp g$. Then there is a
sequence $\epsilon_n\to 0$ for which each $g_n=g+\epsilon_n\chi$ is an
element of $\tCoinX{I;\RR}$. 
\end{Lem}
{\em Proof:} If $g$ is identically zero the result is trivial, so we
assume henceforth that it is not, so $M=\sup |g''|$ is strictly positive.
Suppose the stated result is false, so there
exists an $\epsilon_0>0$
such that $g+\epsilon\chi\notin\tCoinX{I;\RR}$ for all $|\epsilon|\le
\epsilon_0$. Choose $N\in\NN$ sufficiently large that
$(N-1)\sqrt{2\epsilon_0/(MN)}$ exceeds the diameter of the support of
$g$. By hypothesis, for each $|\epsilon|<\epsilon_0$ the function
$g+\epsilon_n\chi$ has a zero of infinite order within its support;
since $\chi\in\tCoinX{I;\RR}$ this zero must lie in the support of $g$
and is therefore a point at which $g'$ vanishes, while $g$ takes the value
$-\epsilon_n$. We may therefore choose
$z_1<z_2<\ldots<z_N$ such that $g'(z_k)=0$ for each $k$ and $g(z_k)$ runs
through the values $\{-k\epsilon_0/N:k=1,\ldots,N\}$ (not necessarily in order). Using
Taylor's theorem with remainder at each $z_k$,
\begin{equation}
\frac{\epsilon_0}{N} \le |g(z_{k+1})-g(z_k)| \le \frac{M}{2}(z_{k+1}-z_{k})^2
\end{equation}
so 
\begin{equation}
z_N-z_1= \sum_{k=1}^{N-1} (z_{k+1}-z_k) \ge
(N-1)\sqrt{\frac{2\epsilon_0}{MN}}
\ge {\rm diam}\,\supp g
\end{equation}
which is a contradiction, since $z_1$ and $z_N$ must belong to the support of
$g$. $\square$

Since $g_n-g=\epsilon_n\chi$, it is clear that $g_n\to g$ and
$g_n^{(k)}\to g^{(k)}$ in $L^2(I)$ as $n\to\infty$. This is the property
required in Sec.~\ref{sect:GA_inertial}.

\begin{acknowledgments}
This work was initiated at the Erwin Schr\"{o}dinger Institute
for Mathematical Physics, Vienna, during a program on Quantum Field
Theory in Curved Spacetime. We are grateful to the organizers of this
program and to the Institute for its hospitality and financial
support. CJF further thanks the
Isaac Newton Institute, Cambridge, for support during the final stages
of the work, and the organisers of the programme on Global Problems in
Mathematical Relativity.
We also thank  L.H.~Ford, A.~Higuchi, B.S.~Kay, J.~Loftin and T.A.~Roman 
for many illuminating discussions, and L.~Osterbrink and C.J.~Smith for
comments on the manuscript. This work was partially supported
by EPSRC Grant GR/R25019/01 to the University of York and by a grant 
from the US Army Research Office through the USMA 
Photonics Research Center. MJP also thanks
the University of York for a grant awarded under its ``Research Funding
for Staff on Fixed Term Contracts'' scheme.
\end{acknowledgments}
\vspace*{.25in}
\noindent The views expressed herein are those of the authors and do not purport
to reflect the position of the United States Military Academy, the
Department of the Army, or the Department of Defense.


\end{document}